\documentclass[aps,pra,twocolumn,superscriptaddress,notitlepage]{revtex4-2}
\usepackage{amsfonts,mathtools,bm,booktabs}
\usepackage{xcolor}
\usepackage[export]{adjustbox}
\usepackage[group-minimum-digits=4,
            per-mode=symbol,
            separate-uncertainty=true,
            range-phrase=--,
            detect-weight=true,
            detect-family=true]{siunitx}
\usepackage[colorlinks=true,
            allcolors=blue,
            breaklinks=true,
            pdftitle={Quantum correlations in molecular cavity optomechanics},
            pdfauthor={Emale Kongkui Berinyuy et al.}]{hyperref}
\usepackage[capitalize,noabbrev]{cleveref}
\Crefname{equation}{Eq.}{Eqs.}
\crefname{equation}{Equation}{Equations}
\begin{document}

\title{Quantum correlations in molecular cavity optomechanics}

\author{E. \surname{Kongkui Berinyuy}}
\email{emale.kongkui@facsciences-uy1.cm}
\affiliation{Department of Physics, Faculty of Science, University of Yaounde I, P.O.Box 812, Yaounde, Cameroon}

\author{D.R. \surname{Kenigoule Massembele}}
\affiliation{Department of Physics, Faculty of Science,
University of Ngaoundere, P.O. Box 454, Ngaoundere, Cameroon}

\author{P. Djorwé}
\email{djorwepp@gmail.com}
\affiliation{Department of Physics, Faculty of Science,
University of Ngaoundere, P.O. Box 454, Ngaoundere, Cameroon}
\affiliation{Stellenbosch Institute for Advanced Study (STIAS), Wallenberg Research Centre at Stellenbosch University, Stellenbosch 7600, South Africa}

\author{R. Altuijri}
\affiliation{Department of Physics, College of Science, Princess Nourah bint Abdulrahman University, P. O. Box 84428, Riyadh 11671, Saudi Arabia}

\author{S. Adbel-Khalek}
\affiliation{Department of Mathematics and Statistics, College of Science, Taif 21944, Saudi Arabia}


\author{A.-H. Abdel-Aty}
\affiliation{Department of Physics, College of Sciences, University of Bisha, Bisha 61922, Saudi Arabia}
\affiliation{Physics Department, Faculty of Science, Al-Azhar University, Assiut 71524, Egypt}

\author{S. G. \surname{Nana Engo}}
\affiliation{Department of Physics, Faculty of Science, University of Yaounde I, P.O.Box 812, Yaounde, Cameroon}

\begin{abstract}
Quantum correlations are interesting resources for modern quantum technologies such as quantum information processing, quantum communication, quantum teleportation, and quantum computation tasks. However, engineering these quantum states turns to be not an easy task. Here, we unveil a theoretical framework for generating and  controlling quantum correlations within a double-cavity molecular optomechanical (McOM) system. Our approach leverages strong interactions between confined optical fields and collective molecular vibrations, creating a versatile environment for exploring robust quantum correlations. Our findings reveal that by judiciously optimizing the coupling strength between the cavity field and the molecular collective mode leads to significant enhancement of entanglement, quantum steering, and quantum discord. We demonstrate that cavity-cavity quantum correlations can be effectively mediated by the molecular collective mode, enabling a unique pathway for inter-cavity quantum connectivity. Moreover, the quantum entanglement generated in our McOM system exhibits robustness against thermal noise, persisting up to temperatures approaching $1000 K$. This strong resilience, qualifies molecular optomechanics as a compelling architecture for scalable, room-temperature quantum information processing and the practical realization of quantum networks. Additionally, the phase-dependent behaviour of quantum discord provides a fundamental basis for developing ultra-sensitive gas sensors, with potential applications in environmental monitoring, medical diagnostics, and industrial safety.
\end{abstract}

\maketitle

\section{Introduction} \label{sec:Intro}

Quantum correlations such as entanglement, steering, and discord form a bottleneck of modern quantum technologies. These non-classical phenomena transcend mere theoretical curiosities, they are the bedrock upon which groundbreaking advances in quantum information processing are built. Whether enabling unconditionally secure communication through quantum cryptography~\cite{Li2003} and facilitating seamless transfer of quantum states via teleportation~\cite{Vaidman1994}, or underpinning the  computational power of quantum algorithms, these correlations are truly required.

While quantum entanglement is widely celebrated as the ultimate expression of inseparable correlations between distant systems~\cite{Lai2022}, Einstein-Podolsky-Rosen (EPR) steering reveals another manifestation of nonlocality. It captures the unique scenario in which one party can remotely influence or \textit{steer} another's quantum state through local measurements – a critical resource for one-sided/nonreciprocal device-independent quantum information protocols~\cite{Kogias2015, Sohail2023}. Expanding beyond entanglement, quantum discord~\cite{Ollivier2001} stands as a profound measure, quantifying the full spectrum of genuinely quantum correlations that can persist even in separable states. Its distinctive ability to unlock quantum advantages in tasks where entanglement is absent has attracted significant attention~\cite{Datta2008,Lanyon2008,Emale2025,Massembele2025}.

Cavity optomechanical (COM) systems have emerged as a leading and exceptionally versatile platform for engineering and controlling these quantum correlations. By  harnessing radiation-pressure coupling between confined optical fields and macroscopic mechanical elements, researchers have extensively theoretically explored and experimentally demonstrated the generation of entanglement~\cite{Xu2010,Agasti2024,Mas2024,Djo2024}, precise control over steering~\cite{Sun_2017}, and sophisticated manipulate discord~\cite{Amazioug2018,Giorda2010,Chakraborty2017}. Building upon this robust foundation, recent investigations have extended the optomechanical paradigm to molecular cavity optomechanics (McOM)~\cite{Huang2024,Schmidt_2024,Zou2024,Xomalis2021,Shalabney2015,Chikkaraddy2022,Roelli2024,Berinyuy2025,Berinyuy2025a}. In these intriguing systems, the collective vibrational modes of molecular ensembles interact directly with confined optical fields. This novel approach promises significant advantages, offering enhanced light-matter nonlinearities and paving the way for a host of unprecedented quantum phenomena, such as optomechanically induced molecular transparency~\cite{Yin2025}, alongside other fascinating developments in molecular polaritons and quantum control~\cite{Xiang2024,Koner2023,Xie2025,Zhang2023,Liu2025,Peng2025,Berinyuy2025,Berinyuy2025a,KoczorBenda2022}.

Despite these exciting developments, a comprehensive characterization of both quantum discord and EPR steering, especially their intricate interplay, within McOM systems remains largely unaddressed. Although the generation of entanglement in molecular optomechanical configurations has received considerable attention~\cite{Huang2024,Schmidt_2024,Berinyuy2025,Berinyuy2025a}, a thorough exploration of more general quantum correlations - those indispensable for quantum networking and asymmetric protocols - are still lacking. This omission is particularly salient given the inherent advantages of molecular systems, notably their  potential for robust operation at room temperatures~\cite{Barzanjeh2021,Degen2017}, a formidable challenge that has long hindered scalable quantum technologies.

It is precisely this significant gap that our current work is aiming to bridge. Therefore, we introduce a rigorous theoretical framework for generating, controlling, and characterizing the full spectrum of quantum correlations, entanglement, steering, and discord within a double-cavity molecular optomechanical system. Our study distinguishes itself through several primary contributions. Firstly, we develop a detailed Hamiltonian model, incorporating collective molecular operators and linearized quantum dynamics, thereby laying a robust theoretical foundation. Secondly, we provide a comprehensive demonstration of EPR steering and Gaussian quantum discord within McOM systems, unveiling their intricate behavior and great potential. Thirdly, we systematically identify the optimal parameter regimes (including detunings $\tilde{\Delta}_a$, $\Delta_c$, coupling $G_j$, and decay $\kappa_j$) essential for maximizing and fine-tuning these diverse types of quantum correlations. Most importantly, our findings strikingly unveil a thermal robustness of our proposed system. It maintains significant quantum correlations at temperatures approaching 1000 K. This stability, surpassing the capabilities of conventional optomechanical platforms, qualifies molecular optomechanics as a compelling architecture for robust, room-temperature quantum information processing, paving the way for practical quantum network implementation.

The rest of this manuscript is structured as follows: \Cref{sec:model} introduces the theoretical model and outlines the derivation of the dynamical equations. \Cref{sec:results} delves into the numerical results and offers a thorough discussion of the underlying quantum correlations. Finally, \Cref{sec:concl} provides concluding remarks and offers insights into promising future research directions.

\section{Theoretical Model and dynamical equations} \label{sec:model}

\begin{figure*}[htp!]
    \centering
    \includegraphics[width=1\linewidth]{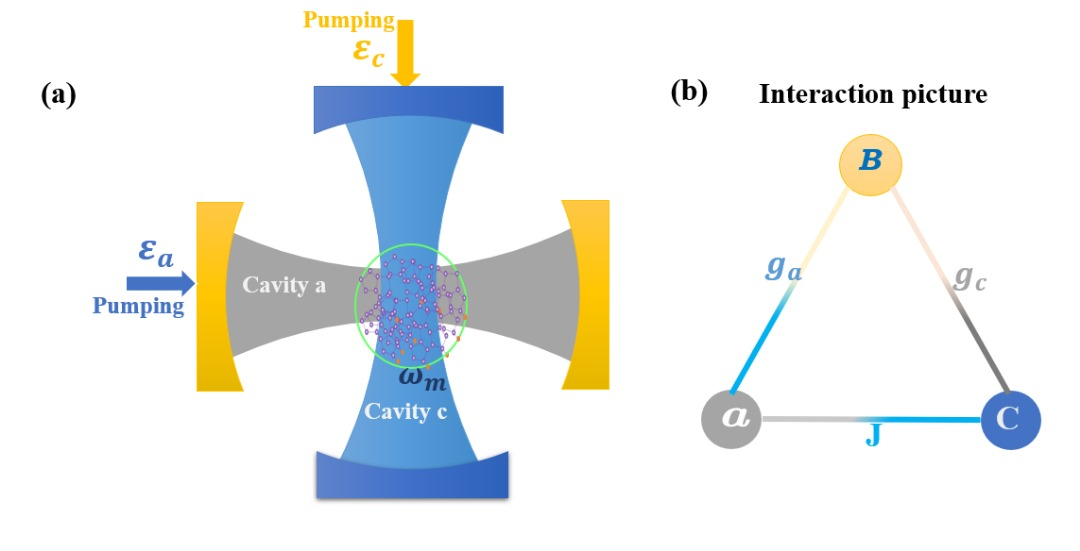}
    \caption{ (a) A molecular optomechanical configuration featuring an ensemble of $N$ molecules strategically placed within two coupled cavities, whose modes are labelled $a$ and $c$. The cavity mode $a$ ($c$) is optomechanically (bilinearly) coupled to the molecular ensemble through the driving filed  $\mathcal{E}_a (\mathcal{E}_c$). (b) An interaction picture highlighting the optomechanical coupling ($g_a, g_c$) between the cavity modes ($a, c$) and the collective molecular vibrational mode ($B$).}
    \label{fig:setup}
\end{figure*}

Our theoretical framework consists of a molecular optomechanical system made of two distinct cavities, labeled $a$ and $c$. Each cavity hosts an ensemble of $N$ molecules, whose collective vibrational modes interact with the confined optical fields. These two cavity field modes interact through a photon tunneling process characterized by a coupling strength $J$. One cavity is driven by a strong driving field (mode $a$ with amplitude $\mathcal{E}_a$) and is optomechanically coupled to the molecules, while the other cavity (mode $c$ with amplitude $\mathcal{E}_c$) is driven via a weak pumping field that leads to a bilinear interaction with the molecules. Our benchmark model is sketched in \Cref{fig:setup}. The collective vibrational motion of these molecules is modeled as a quantum harmonic oscillator. The total Hamiltonian of the system, expressed in a frame rotating at the driving frequency $\omega_\ell$, is given by:
\begin{equation}\label{eq:1}
\begin{aligned}
H = & \Delta_a a^\dagger a + \Delta_c c^\dagger c  + \sum_{j=1}^N \omega_m b_j^\dagger b_j + \sum_{j=1}^N g_a a^\dagger a(b_j^\dagger + b_j) 
     \\&+ \sum_{j=1}^N g_c(c^\dagger+c)(b_j^\dagger + b_j)+J(a^\dagger c+ac^\dagger) + i\mathcal{E}_a(a^\dagger - a)\\& + i\mathcal{E}_c(c^\dagger - c),
\end{aligned}
\end{equation}
where $a (a^\dagger)$, $c (c^\dagger)$, and $b_j (b_j^\dagger)$ are the annihilation (creation) operators for the cavity modes $a$ and $c$, and the $j$-th individual molecular vibrational mode, respectively. These operators satisfy the canonical commutation relations, e.g., $[a, a^\dagger] = 1$. The parameter $\omega_m$ denotes the resonance frequency of the molecular vibrational mode. The terms $\Delta_a = \omega_a - \omega_\ell$ and $\Delta_c = \omega_c - \omega_\ell$ represent the frequency detunings of the effective cavity modes $a$ and $c$ relative to the driving field frequency. The first two terms capture the free Hamiltonians of the two cavity modes. The fourth and fifth terms describe the optomechanical coupling  between cavity mode $a$ with the molecules, and the bilinear coupling  between mode $c$ and molecular vibration modes, respectively. These two couplings are captured via the coupling strengths $g_a$ and $g_c$. Finally, the last two terms account for the external driving of the cavity modes by the pumping fields.

To simplify the description of the molecular ensemble, especially when each molecule interacts identically with the cavity fields, we introduce a collective molecular vibrational operator $B$,
\begin{equation}\label{eq:2}
B = \frac{1}{\sqrt{N}}\sum_{j=1}^N b_j.
\end{equation}
This collective mode $B$ is a bosonic operator that satisfies the commutation relation $[B,B^\dagger]=1$, which arises naturally from the sum of the individual bosonic modes $N$. With this collective operator, our Hamiltonian can be compactly rewritten as,
\begin{equation}\label{eq:3}
\begin{aligned}
H = & \Delta_a a^\dagger a + \Delta_c c^\dagger c  + \omega_m B^\dagger B + G_aa^\dagger a(B^\dagger + B) 
    +\\& G_c(c^\dagger + c)(B^\dagger + B)+J(a^\dagger c+ac^\dagger) + i\mathcal{E}_a(a^\dagger - a)\\& + i\mathcal{E}_c(c^\dagger - c),
\end{aligned}
\end{equation}
where $G_a = g_a\sqrt{N}$ and $G_c = g_c\sqrt{N}$ represent the collective optomechanical coupling strengths, reflecting the significant enhancement achievable due to the large number of molecules.

\subsection{Quantum Langevin Equations and linearization}\label{sec:qle} 

The temporal evolution of our system is described by the following set of Quantum Langevin Equations (QLEs), which incorporate dampings and quantum noises inherent to open quantum systems:
\begin{equation}\label{eq:4}
\begin{aligned}
\dot{a}=&-(i\Delta_{a}+\kappa_a)a-G_aa(B^\dagger+B)-iJc+\mathcal{E}_a+\sqrt{2\kappa_a}a^{in},\\
\dot{c}=&-(i\Delta_c+\kappa_c)c-iG_c(B+B^\dagger)-iJa+\mathcal{E}_c+\sqrt{2\kappa_c}c^{in},\\
\dot{B}=&-(i\omega_m+\gamma_m)B+iG_aa^\dagger a+G_c(c^\dagger +c)+\sqrt{2\gamma_m}B^{in},
\end{aligned}
\end{equation}
where $\kappa_a$ and $\kappa_c$ are the decay rates of cavity modes $a$ and $c$, respectively, and $\gamma_m$ is the mechanical damping rate. The operators $a^{\text{in}}$, $c^{\text{in}}$, and $B^{\text{in}}$ represent the corresponding input noise operators. Notably, $B^{\text{in}}$ is the collective noise operator for the molecular ensemble, defined consistently with $B$ as $B^{\text{in}} = \frac{1}{\sqrt{N}}\sum_{j=1}^N b^{\text{in}}_j$. These noise operators have zero mean values and their correlation functions are characterized by,
\begin{equation}\label{eq:5}
\begin{aligned}
&\langle a^{\text{in}}(t)a^{\text{in}\dagger}(t^\prime)\rangle = \delta(t-t^\prime), 
\langle c^{\text{in}}(t)c^{\text{in}\dagger}(t^\prime)\rangle = \delta(t-t^\prime), \\
&\langle B^{\text{in}}(t)B^{\text{in}\dagger}(t^\prime)\rangle = (n_{\text{th}} + 1)\delta(t-t^\prime),\\& 
\langle B^{\text{in}\dagger}(t)B^{\text{in}}(t^\prime)\rangle = n_{\text{th}}\delta(t-t^\prime),
\end{aligned}
\end{equation}
where $n_{\text{th}} = \left[\exp\left(\hbar\omega_m/k_B T\right) - 1\right]^{-1}$ is the thermal phonon number for the mechanical mode at temperature $T$, and $k_B$ is the Boltzmann constant. It is important to note that for the optical cavity modes, thermal excitation is typically negligible in the optical frequency band, allowing their corresponding thermal phonon numbers to be set to zero.

To analyze the system's quantum dynamics in the continuous-variable regime, we employ a standard linearization procedure, assuming a strong classical drive of the cavities. This allows us to split each operator $\mathcal{O}$ into its classical mean value $\langle\mathcal{O}\rangle$ and a small quantum fluctuation $\delta\mathcal{O}$, i.e., $\mathcal{O} = \langle\mathcal{O}\rangle + \delta\mathcal{O}$, where $\mathcal{O} \in \{a, c, B\}$. For long-term steady-state behavior, these operators settle to their mean values ($\alpha_a = \langle a \rangle$, $\alpha_c = \langle c \rangle$, and $\beta = \langle B \rangle$), which are governed by the following equations,
\begin{equation}\label{eq:steady_state_means}
\begin{aligned}
(i\Delta_a + \kappa_a)\alpha_a +iG_a\alpha_a(\beta^* + \beta)+iJ\alpha_c - \mathcal{E}_a &= 0, \\
(i\Delta_c + \kappa_c)\alpha_c + iG_c(\beta^* + \beta)+iJ\alpha_a - \mathcal{E}_c &= 0, \\
(i\omega_m + \gamma_m)\beta - iG_a|\alpha_a|^2 - G_c(\alpha_c^* + \alpha_c) &= 0.
\end{aligned}
\end{equation}
In this regime, the system's dynamics are well-described by the linearized QLEs for the fluctuation operators,
\begin{equation}\label{eq:6}
\begin{aligned}
\delta\dot{a}&=-(i\Delta_a+\kappa_a)\delta a-\tilde{G}_ae^{i\theta_1}(B^\dagger+B)-iJ\delta c+\sqrt{2\kappa_a}a^{in}\\
\delta\dot{c}&=-(i{\Delta}_c+\kappa_c)\delta c-iG_ce^{i\theta_2}(\delta B+\delta B^\dagger)-iJ\delta a+\sqrt{2\kappa_c}c^{in},\\
\delta\dot{B}&=-(i\omega_m+\gamma_m)\delta B-ie^{i\theta_1}(\tilde{G}_a\delta a^\dagger+G_c\delta c^\dagger)-\\&ie^{-i\theta_1}(\tilde{G}^\ast_a\delta a+G^\ast_c\delta c)+\sqrt{2\gamma_m}B^{in}
\end{aligned}
\end{equation}
where $\tilde{\Delta}_a = \Delta_a + G_a(\beta^* + \beta)$ is the effective detuning, incorporating the mean displacement of the molecular mode and $\theta_j$ is the phase of the intracavity field. Similarly, $\tilde{G}_a = G_a|\alpha_a|$  represents the linearized collective optomechanical coupling strength. We choose a phase reference such that $\alpha_a$ and $\alpha_c$ are real numbers.

It is important to note that the individual intracavity phases $\theta_j$ can be absorbed into a redefinition of the fluctuation operators $\delta a_j$, so that only the relative phase $\theta = \theta_1 - \theta_2$ has physical significance. From \cref{eq:6}, we can derive an effective linearized Hamiltonian under the rotating wave approximation (RWA). This Hamiltonian reveals the rich interplay between the cavity modes and the collective molecular vibration,
\begin{equation}\label{eq:lin}
\begin{aligned}
    H_{\text{lin}} = & \tilde{\Delta}_a\delta a^\dagger\delta a + \Delta_c \delta c^\dagger\delta c  + \omega_m \delta B^\dagger \delta B + \tilde{G}_a(\delta a^\dagger + \delta a)\\& (\delta B^\dagger + \delta B) + G_c(\delta c^\dagger + \delta c)(\delta B^\dagger + \delta B)+\\&J(e^{i\theta}\delta a^\dagger\delta c+e^{-i\theta}\delta c^\dagger\delta a).
\end{aligned}
\end{equation}
This linearized Hamiltonian encompasses a mix of resonant and non-resonant interaction terms. When the cavity $a$ is detuned to the anti-Stokes sideband (i.e., $\tilde{\Delta}_a = \omega_m$, commonly known as red detuning), the interaction within RWA takes the form $H_{\text{int}} = \tilde{G}_a(\delta a^\dagger\delta B + \delta a\delta B^\dagger)$. This represents a beam-splitter-like interaction, which facilitates coherent energy exchange between the cavity mode $a$ and the molecular collective mode. Furthermore, in a thermally populated environment, applying red-detuned driving effectively cools the molecular collective mode by extracting vibrational energy, thereby creating conditions leading quantum correlations. Conversely, if the cavity $c$ is detuned to the Stokes sideband (that is, $\Delta_c = -\omega_m$, commonly known as blue detuning), the Hamiltonian interaction becomes $H_{\text{int}} = G_c(\delta c\delta B + \delta c^\dagger\delta B^\dagger)$. This term corresponds to a two-mode squeezing interaction, which directly generates quantum correlations between the cavity mode $c$ and the molecular collective mode.

\subsection{Quadrature Operators and covariance matrix formalism}\label{sec:cm_formalism}

To fully characterize the Gaussian quantum state of the system, we introduce  quadrature operators, which correspond to position and momentum observables,
\begin{equation}
\begin{aligned}
&\delta X_a = \frac{( \delta a + \delta a^{\dagger})}{\sqrt2}, \quad \delta Y_a = \frac{( \delta a - \delta a^{\dagger})}{i\sqrt2}, \\
&\delta X_c = \frac{( \delta c + \delta c^{\dagger})}{\sqrt2}, \quad \delta Y_c = \frac{( \delta c - \delta c^{\dagger})}{i\sqrt2}, \\
&\delta Q = \frac{(\delta B + \delta B^{\dagger})}{\sqrt2}, \quad \delta P = \frac{(\delta B - \delta B^{\dagger})}{i\sqrt2}.
\end{aligned}
\end{equation}
The corresponding input noise operators for these quadratures are,
\begin{equation}
\begin{aligned}
X_a^{\text{in}} &= \frac{(a^{\text{in}} + a^{\text{in}\dagger})}{\sqrt2}, \quad Y_a^{\text{in}} = \frac{(a^{\text{in}} - a^{\text{in}\dagger})}{i\sqrt2}, \\
X_c^{\text{in}} &= \frac{(c^{\text{in}} + c^{\text{in}\dagger})}{\sqrt2}, \quad Y_c^{\text{in}} = \frac{(c^{\text{in}} - c^{\text{in}\dagger})}{i\sqrt2}, \\
Q^{\text{in}} &= \frac{(B^{\text{in}} + B^{\text{in}\dagger})}{\sqrt2}, \quad P^{\text{in}} = \frac{(B^{\text{in}} - B^{\text{in}\dagger})}{i\sqrt2}.
\end{aligned}
\end{equation}
To facilitate analysis, we cast these linearized QLEs from \cref{eq:6} into a compact matrix form using the defined quadrature operators. This is the primary method used for our numerical simulations across all parameter regimes,
\begin{equation}\label{eq:7}
\dot{\bm{u}}(t) = A\bm{u}(t) + \bm{n}(t),
\end{equation} 
where $\bm{u}^\top(t) = \left( \delta X_a(t), \delta Y_a(t), \delta X_c(t), \delta Y_c(t), \delta Q(t), \delta P(t) \right)$ is the vector of fluctuation quadrature operators, and $\bm{n}^\top(t) = \left( \sqrt{2\kappa_a}X_a^{\text{in}}, \sqrt{2\kappa_a}Y_a^{\text{in}}, \sqrt{2\kappa_c}X_c^{\text{in}}, \sqrt{2\kappa_c}Y_c^{\text{in}}, \sqrt{2\gamma_m}Q^{\text{in}}, \sqrt{2\gamma_m}P^{\text{in}} \right)$ is the corresponding noise vector. The drift matrix $A$ is a $6\times 6$ matrix defined as,
\begin{equation}\label{eq:drift_matrix_A}
\mathcal{A}=
\begin{pmatrix}
-\kappa_a & \tilde{\Delta}_a & J \sin \theta & J \cos \theta & 0 & 0 \\
-\tilde{\Delta}_a & -\kappa_a & -J \cos \theta & J \sin \theta & -2 \tilde{G}_a & 0 \\
- J \sin \theta & J \cos \theta & -\kappa_c & \Delta_c & 0 & 0 \\
- J \cos \theta & - J \sin \theta & -\Delta_c & -\kappa_c & -2 G_c & 0 \\
0 & 0 & 0 & 0 & -\gamma_m & \omega_m \\
- 2 \tilde{G}_a & 0 & -2 G_c & 0 & -\omega_m & -\gamma_m
\end{pmatrix}.
\end{equation}

For the system to be dynamically stable, all real parts of the eigenvalues of the drift matrix $A$ must be negative. This essential stability condition is rigorously established using the Routh-Hurwitz criterion~\cite{DeJesus1987}. Given the Gaussian nature of quantum noises and the linearity of QLEs, the quantum state of the system can be fully characterized by a $6 \times 6$ covariance matrix (CM), denoted $V$. This CM is obtained by solving the continuous-variable Lyapunov equation~\cite{Gardiner2000,Vidal2002},
\begin{equation}\label{eq:lyapunov}
A\bm{V} + \bm{V}A^\top = -\bm{D},
\end{equation}
where $\bm{V}_{lk} = \frac12\left\{ \langle \bm{u}_\ell(t)\bm{u}_k(t^{\prime}) \rangle + \langle \bm{u}_k(t^{\prime})\bm{u}_\ell(t) \rangle \right\}$ represents the symmetric correlation functions of the fluctuation operators in the steady state. The diffusion matrix $\bm{D}$, which accounts for the influence of the noise operators, is given by,
\begin{equation}\label{eq:diffusion_matrix_D}
\bm{D} = \text{diag}\left[ \kappa_a, \kappa_a, \kappa_c, \kappa_c, \gamma_m(2n_{\text{th}} + 1), \gamma_m(2n_{\text{th}} + 1) \right].
\end{equation}

\subsection{Quantification of Bipartite Entanglement}\label{sec:entanglement}

To quantify the bipartite entanglement present in our system, we employ the logarithmic negativity, $E_n$, a well-established entanglement monotone for Gaussian states~\cite{Plenio2005}. Its mathematical expression is given by,
\begin{equation}\label{eq:log_neg_En}
E_n = \max\left[0, -\ln(2\Gamma)\right],
\end{equation}
where $\Gamma$ is a function of the covariance matrix elements, specifically:
\begin{equation}\label{eq:log_neg_Gamma}
\Gamma \equiv \frac{1}{\sqrt2} \sqrt{\Sigma(V) - \sqrt{\Sigma(V)^2 - 4\det(V)}},
\end{equation}
with $\Sigma(V) = \det(\Psi_1) + \det(\Psi_2) - 2\det(\Psi_3)$. For any given bipartite subsystem (e.g., cavity $a$ and cavity $c$, or cavity $a$ and the collective molecular mode $B$), the relevant $4\times4$ submatrix $V_{\text{sub}}$ extracted from the full $6\times6$ CM takes the canonical block form:
\begin{equation}\label{eq:V_sub_block_form}
V_{\text{sub}} =
\begin{pmatrix}
\Psi_1 & \Psi_3 \\
\Psi_3^T & \Psi_2 \\
\end{pmatrix},
\end{equation}
where $\Psi_1$ and $\Psi_2$ are $2\times2$ symmetric matrices representing the local correlations of each subsystem, and $\Psi_3$ is a $2\times2$ matrix encoding the inter-subsystem correlations.

\subsection{Einstein-Podolsky-Rosen (EPR) steering}\label{sec:steering}

Beyond entanglement, EPR steering offers a distinct and asymmetric measure of quantum correlation, providing deeper insights into the non-classical behavior of quantum systems. Steering describes the scenario where one party (Alice) can remotely prepare the quantum state of another party (Bob) through her local measurements. The steerability from mode $a$ to mode $c$ ($\mathcal{G}_{a\to c}$) and vice versa ($\mathcal{G}_{c\to a}$) for Gaussian states can be quantified using the negativity-based measure introduced by \textcite{Kogias2015},
\begin{align}\label{eq:steering_Gac}
&\mathcal{G}_{a\to c} = \max\left[0, \frac12\ln\frac{\det V_a}{4\det V_{ac}}\right], \\ \label{eq:steering_Gca}
&\mathcal{G}_{c\to a} = \max\left[0, \frac12\ln\frac{\det V_c}{4\det V_{ac}}\right].
\end{align}
Here, $V_a$ and $V_c$ are the covariance matrices $2\times2$ of the modes $a$ and $c$, respectively, while $V_{ac}$ is the covariance matrix $4\times4$ of the combined modes $a$ and $c$. It is important to note that while every steerable quantum state is necessarily entangled (non-separable), the converse is not true; not all entangled states exhibit steerability. This distinction gives rise to different steering phenomena, i.e., no-way steering ($\mathcal{G}_{a\to c} = \mathcal{G}_{c\to a} = 0$), one-way steering (e.g., $\mathcal{G}_{a\to c} > 0$ and $\mathcal{G}_{c\to a} = 0$), and two-way steering ($\mathcal{G}_{a\to c} > 0$ and $\mathcal{G}_{c\to a} > 0$).

\subsection{Gaussian Quantum Discord (GQD)}\label{sec:discord}

To provide a more complete characterization of quantum correlations, we investigate the Gaussian Quantum Discord (GQD). GQD quantifies the total non-classical correlations, including those present even in separable states, making it a powerful measure for various quantum information tasks where entanglement might be absent~\cite{Ollivier2001,Giorda2010}. For a given bipartite Gaussian subsystem $l$, characterized by its $4\times4$ covariance matrix $V_{\text{sub}}$, the GQD is given by~\cite{Giorda2010},
\begin{equation}\label{eq:GQD_DGl}
D_G^\ell = f(\sqrt{I_1^\ell}) - f(v^\ell_-) - f(v^\ell_+) + f(\sqrt{\mathcal{W}^\ell}),
\end{equation}
where the function $f(x)$ is defined as,
\begin{equation}\label{eq:GQD_f_x}
f(x) = \left(x + \frac12\right)\ln\left(x + \frac12\right) - \left(x - \frac12\right)\ln\left(x - \frac12\right).
\end{equation}
The symplectic eigenvalues $v^\ell_-$ and $v^\ell_+$ of the subsystem's covariance matrix are obtained from,
\begin{equation}\label{eq:GQD_v_pm}
v^\ell_{\pm} \equiv \frac{1}{\sqrt2} \left\{ \Sigma(V^\ell_{\text{sub}}) \pm \left[ \Sigma(V^\ell_{\text{sub}})^2 - 4I^\ell_4 \right]^{1/2} \right\}^{1/2},
\end{equation}
with $\Sigma(V_{\text{sub}}) = I^\ell_1 + I^\ell_2 + 2I^\ell_3$. Here, $I^\ell_1 = \det(\Psi_1)$, $I^\ell_2 = \det(\Psi_2)$, $I^\ell_3 = \det(\Psi_3)$, and $I^\ell_4 = \det(V_{\text{sub}})$. The term $\mathcal{W}^\ell$ is an essential quantity that characterizes the quantumness of correlations and is given by~\cite{Chakraborty2017},
\begin{widetext}
\begin{equation}\label{eq:GQD_W_ell}
\mathcal{W}^\ell =
\begin{cases}
\left( \frac{2|I^\ell_3| + \sqrt{4I_3^{l2} + (4I^\ell_1 - 1)(4I^\ell_4 - I^\ell_2)}}{4I^\ell_1 - 1} \right)^2 & \text{if} \quad \frac{4(I^\ell_1I^\ell_2 - I^\ell_4)^2}{(I^\ell_2 + 4I^\ell_4)(1 + 4I^\ell_1)I^{l2}_3} \leq 1, \\
\frac{I^\ell_1I^\ell_2 + I^\ell_4 - I^{l2}_3 - \sqrt{(I^\ell_1I^\ell_2 + I^\ell_4 - I^{l2}_3)^2 - 4I^\ell_1I^\ell_2I^\ell_4}}{2I^\ell_1} & \text{otherwise}.
\end{cases}
\end{equation}
\end{widetext}
A key property is that a Gaussian state is entangled if and only if its smallest symplectic eigenvalue $v_-$ is less than $1/2$. This implies that while $0 \leq D^\ell_G \leq 1$ can correspond to separable or entangled states, a value of $D^\ell_G > 1$ definitively indicates an entangled state~\cite{Giorda2010}.

\section{Results}\label{sec:results}

In this section, we provide the comprehensive findings of our investigation, with a particular emphasis on the behavior of bipartite entanglement, EPR steering, and quantum discord within our molecular optomechanical system. To ensure the robustness and experimental relevance of our simulations, we have meticulously selected parameters inspired by recent experimental advancements in molecular cavity optomechanics and related optomechanical systems~\cite{Schmidt_2024,Xiao2012,Zou2024,KoczorBenda2022,Xomalis2021,Shalabney2015,Chikkaraddy2022,Roelli2024}. These parameters are set as follows: the molecular vibrational frequency $\omega_m/2\pi = \SI{30}{\tera\hertz}$, cavity decay rates $\kappa_c/2\pi = \SI{0.5}{\tera\hertz}$ and $\kappa_a/2\pi = \SI{30}{\tera\hertz}$, normalized driving field amplitude $\mathcal{E}/\omega_m = 16$, mechanical damping rate $\gamma_m/2\pi = \SI{0.16}{\tera\hertz}$, individual molecular coupling strengths $g_c/2\pi = \SI{0.1}{\giga\hertz}$ and $g_a/2\pi = \SI{0.08}{\giga\hertz}$, an ensemble size $N = \num{e6}$ molecules, and an ambient temperature $T = \SI{210}{\kelvin}$.
\subsection{Bipartite entanglement, steering, quantum discord analysis and phase difference $\theta$}
\begin{figure*}[tbh]
	\centering
	\includegraphics[width=5.5cm]{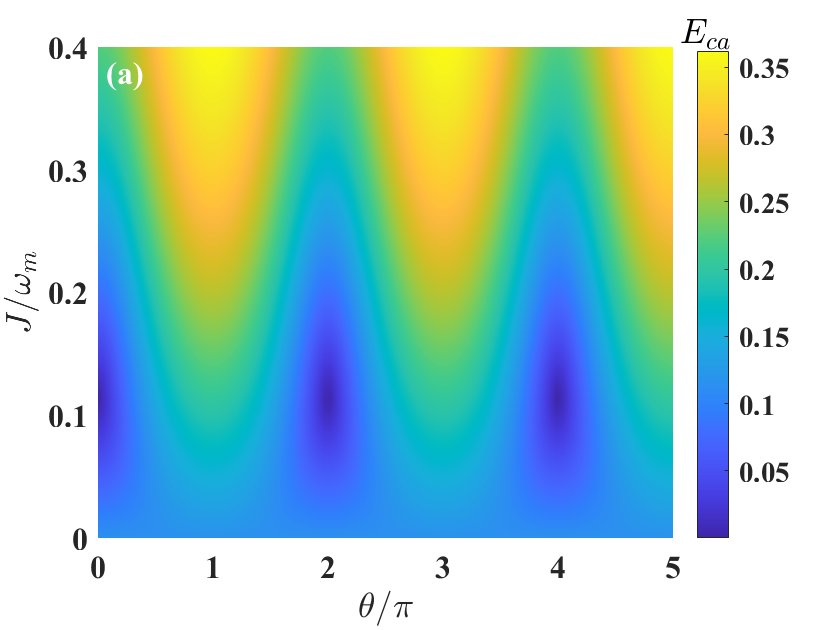}
	\includegraphics[width=5.5cm]{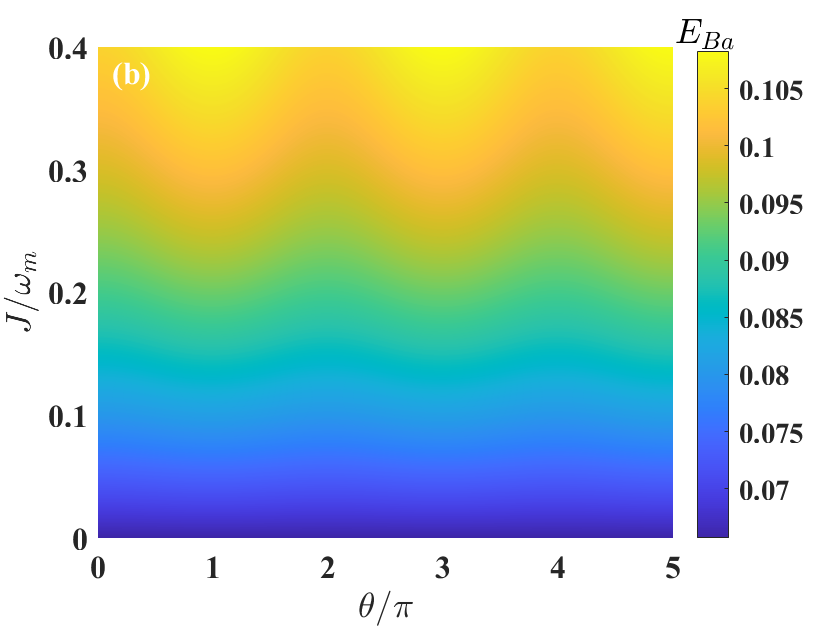}
	\includegraphics[width=5.5cm]{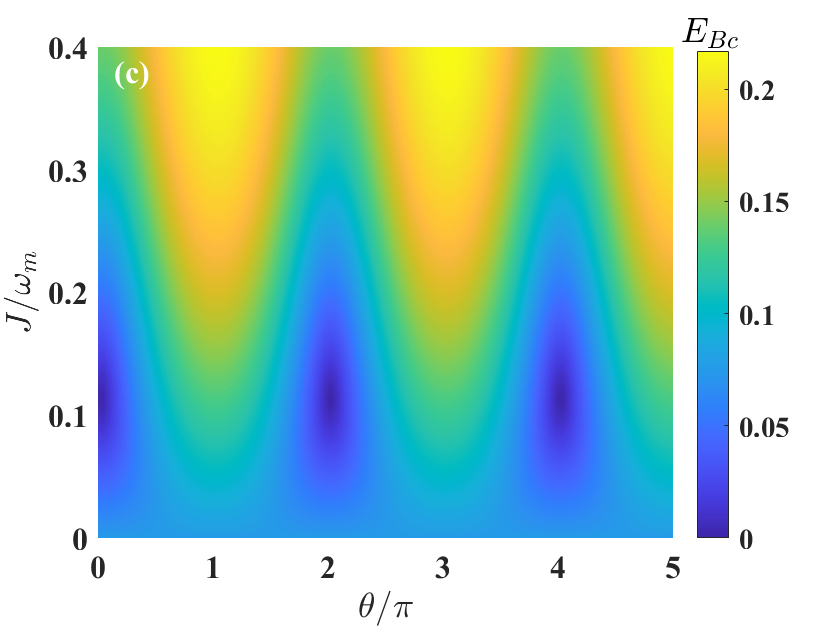}
	\caption{(a) Density plot of bipartite entanglement for photon-photon modes ($E_{ca}$) for  $\kappa_j/\omega_m = \num{0.003}$, and $G_j/\omega_m = \num{0.004}$. (b) Entanglement for vibration-photon modes ($E_{Ba}$) for $\tilde{\Delta}_a/\omega_m = 0.95$, $G_j/\omega_m =0.004$, $\kappa_c/\omega_m = \num{0.003}$, $\kappa_a = \kappa_c$ and  ${\Delta}_c/\omega_m = -0.4$. (c) Entanglement for vibration-photon modes ($E_{Bc}$) for  $\kappa_j/\omega_m = \num{0.05}$, and $G_j/\omega_m = \num{0.005}$ as a function of phase $\theta/\pi$ and $J/\omega_m$. Common parameters for all subplots: $\mathcal{E}/\omega_m = 16$, $\gamma_m/\omega_m = \num{0.005}$, $g_c/\omega_m = \num{3.3e-6}$, $g_a/\omega_m = \num{2.66e-6}$, $\kappa_c/\omega_m = \num{0.0166}$, $\kappa_a = \kappa_c$, $\tilde{\Delta}_a/\omega_m = 1$ and ${\Delta}_c/\omega_m = -1$, and $T = \SI{210}{\kelvin}$.}
	\label{fig:fig2}
\end{figure*}

\begin{figure*}[tbh]
	\centering
	\includegraphics[width=5.5cm]{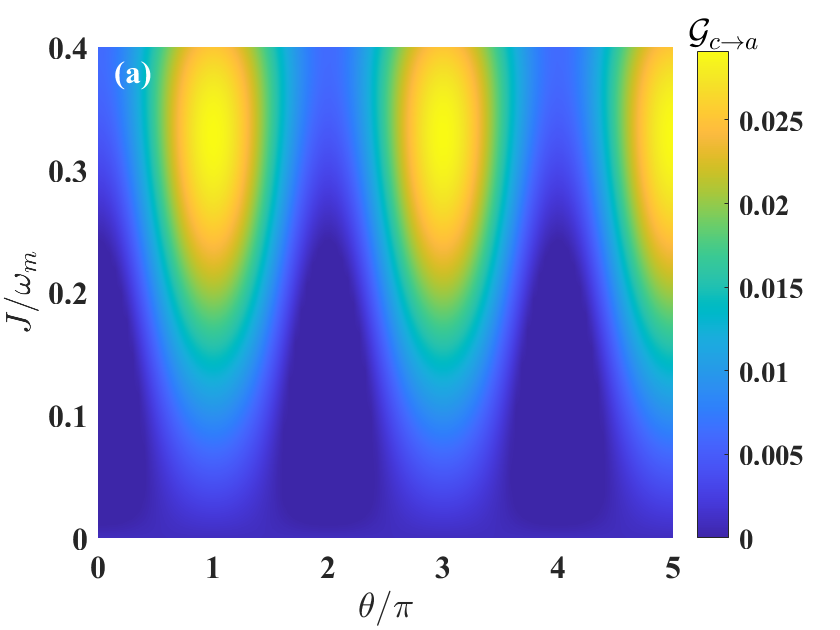}
	\includegraphics[width=5.5cm]{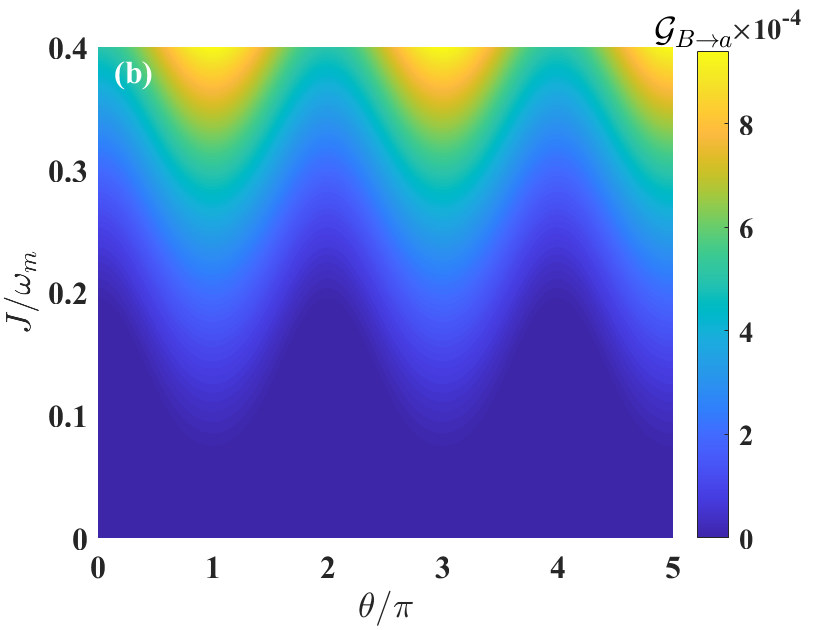}
	\includegraphics[width=5.5cm]{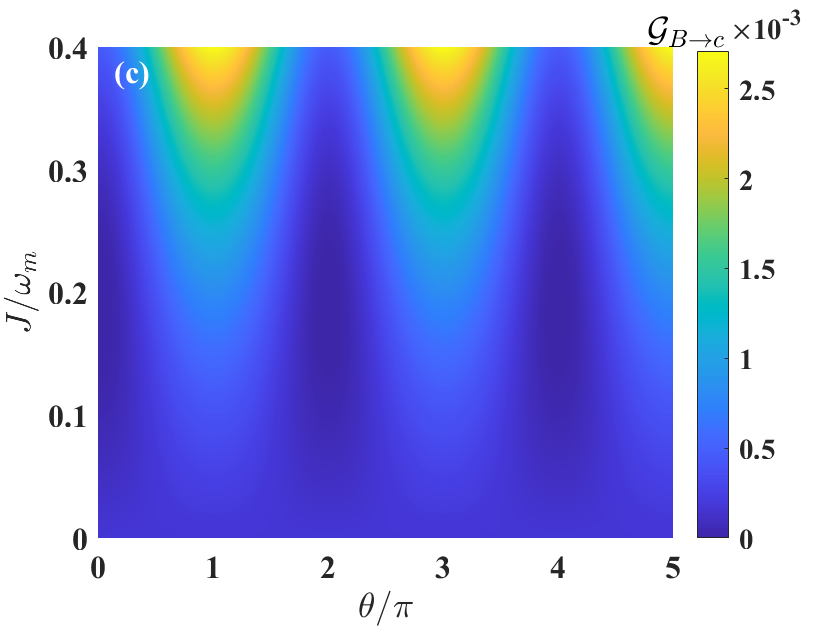}
	\caption{(a) Density plot of one-way steering $\mathcal{G}_{c\to a}$  for  $\kappa_j/\omega_m = \num{0.003}$, and $G_j/\omega_m = \num{0.003}$. (b) $\mathcal{G}_{B\to a}$ for $\tilde{\Delta}_a/\omega_m = 1.5$, ${\Delta}_c/\omega_m = -0.5$, $\kappa_j/\omega_m = \num{0.005}$, and $G_j/\omega_m = \num{0.005}$. (c) $\mathcal{G}_{B\to c}$ for $\tilde{\Delta}_a/\omega_m = 1.5$, ${\Delta}_c/\omega_m = -0.5$, $\kappa_j/\omega_m = \num{0.05}$, and $G_j/\omega_m = \num{0.005}$. Common parameters for all subplots: $\mathcal{E}/\omega_m = 16$, $\gamma_m/\omega_m = \num{0.005}$, $g_c/\omega_m = \num{3.3e-6}$, $g_a/\omega_m = \num{2.66e-6}$, $N = \num{e6}$, $\kappa_c/\omega_m = \num{0.0166}$, $\kappa_a = \kappa_c$, and $T = \SI{210}{\kelvin}$.}
	\label{fig:fig3}
\end{figure*}
\begin{figure*}[tbh]
	\centering
	\includegraphics[width=5.5cm]{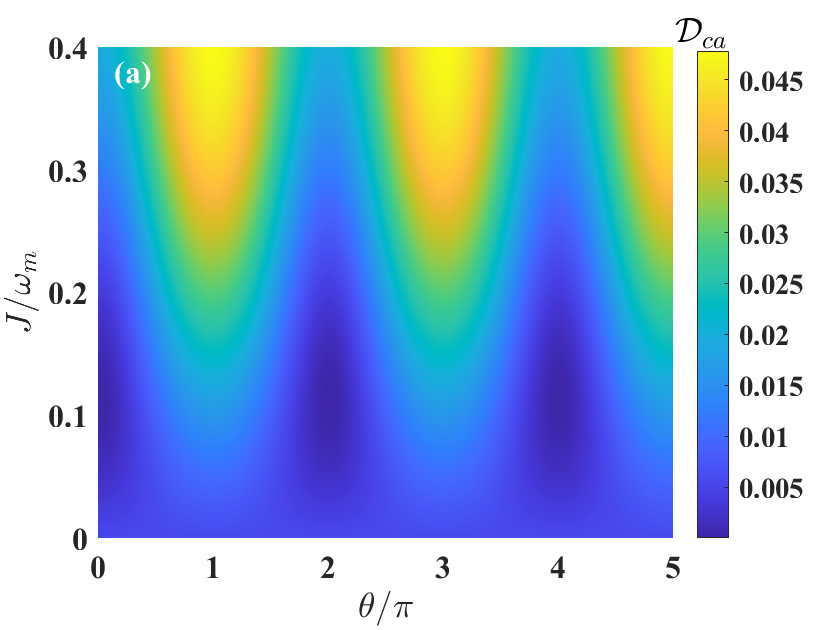}
	\includegraphics[width=5.5cm]{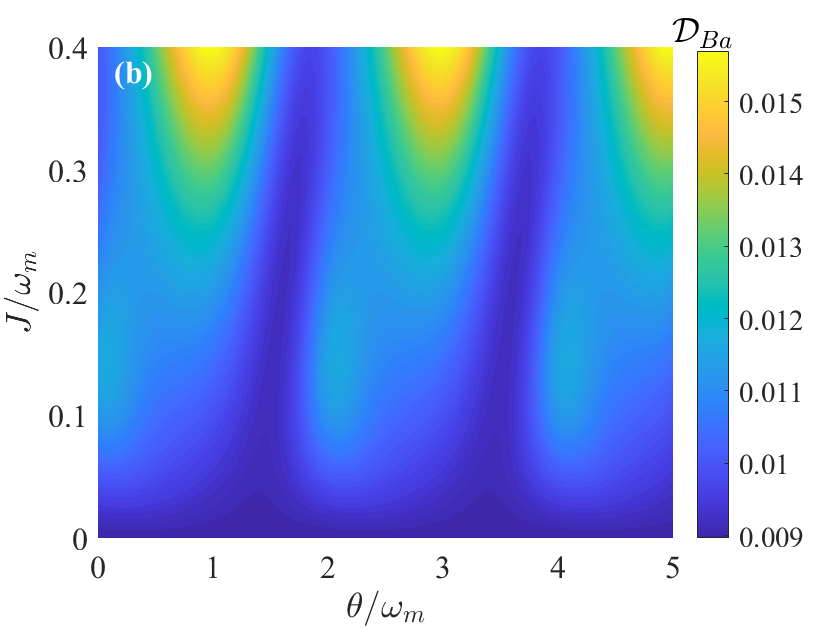}
	\includegraphics[width=5.5cm]{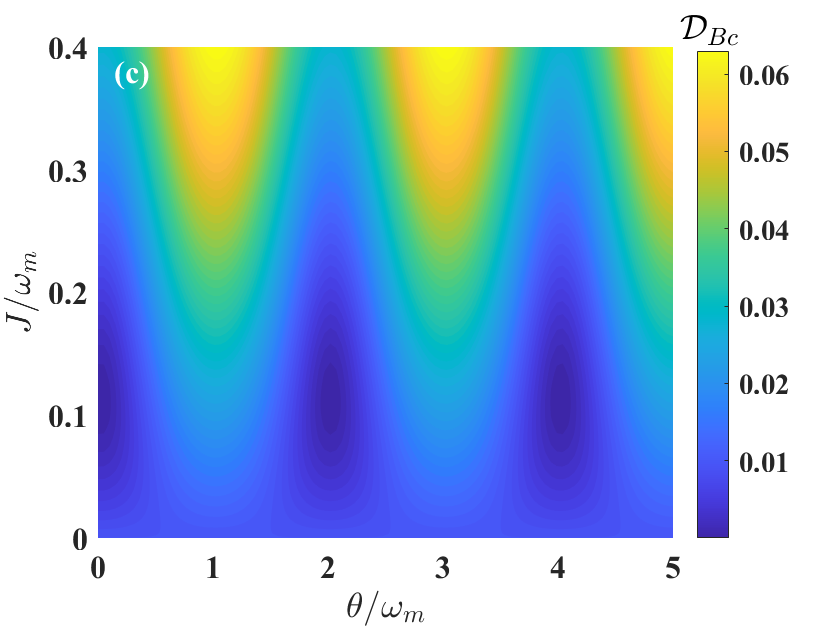}
	\caption{(a) Density plot of quantum discord $\mathcal{D}_{ca}$ between cavity modes $a$ and $c$ for  $\kappa_j/\omega_m = \num{0.05}$, and $G_j/\omega_m = \num{0.005}$. (b) Quantum discord $\mathcal{D}_{Ba}$ between the molecular collective mode $B$ and cavity mode $a$ for  $\kappa_j/\omega_m = \num{0.05}$, and $G_j/\omega_m = \num{0.005}$. (c) Quantum discord $\mathcal{D}_{Bc}$ between the molecular collective mode $B$ and cavity mode $c$ for  $\kappa_j/\omega_m = \num{0.05}$, and $G_j/\omega_m = \num{0.005}$, as a function of phase $\theta/\pi$ and $J/\omega_m$. Common parameters for all subplots: $\mathcal{E}/\omega_m = 16$, $\gamma_m/\omega_m = \num{0.005}$, $g_c/\omega_m = \num{3.3e-6}$, $g_a/\omega_m = \num{2.66e-6}$, $N = \num{e6}$, $\tilde{\Delta}_a/\omega_m = 1$ and ${\Delta}_c/\omega_m = -1$, $\tilde{\Delta}_a/\omega_m = 1$ and ${\Delta}_c/\omega_m = -1$, and $T = \SI{210}{\kelvin}$.}
	\label{fig:fig4}
\end{figure*}
\Cref{fig:fig2} shows the density plot of bipartite entanglement between subsystems as a function of cavity-cavity coupling strengths $J$ and phase difference $\theta$. It is clearly observed that the entanglement is modulated by the phase $\theta$. This entanglement is optimal when $\theta=m\pi$ for $m\in\mathbb{N}$ and drops when $\theta=n\frac{\pi}{2}$ when $n$ is an odd integer. Interestingly, our findings reveal a striking reversal of the commonly observed phase dependence of entanglement. While earlier studies report optimal entanglement at $\theta=n\frac{\pi}{2}$~\cite{Mas2024}, our molecular optomechanical system shows maximum entanglement at $\theta=m\pi$. This unexpected behaviour underscores the distinctive dynamics introduced by the molecular component, suggesting that phase-dependent control in hybrid systems may differ fundamentally from conventional cavity-based setups.

Furthermore, we observe that the one-way steering in our molecular optomechanical system also reaches its optimal value when the phase difference satisfies $\theta=m\pi$ for $m\in\mathbb{N}$ (see \Cref{fig:fig3}). This coincides with the phase values at which bipartite entanglement is maximized, indicating a consistent and robust phase-dependent quantum correlation behaviour. \Cref{fig:fig4} reveal that quantum discord in our molecular optomechanical system exhibits the same phase dependence, reaching its maximum at $\theta=m\pi$, consistent with the behaviours observed for both entanglement and one-way steering. This unified phase sensitivity across different quantum correlation measures entanglement, steering, and discord highlights the distinctive nature of the coupling dynamics in our system. These findings collectively point to a new regime of phase control in quantum hybrid systems, where both entanglement and quantum steering can be co-optimized through discrete phase settings. Such phase-engineered behaviours may pave the way for novel applications in directional quantum communication and sensing using molecular-scale platforms.

As it can be seen from the figures \Cref{fig:fig2} to \Cref{fig:fig4}, the bipartite entanglement, steering and quantum discord are enhanced as the coupling strength between the two cavities $J$ increases for $\theta=m\pi$. This enhancement of entanglement, steering, and discord with increasing coupling strength between the two cavities can be understood as a consequence of stronger coherent interactions between modes, facilitating efficient quantum correlations. When the cavities are more strongly coupled, photons and excitations can transfer more readily between them, enabling better correlation of their quantum states.

\subsection{Bipartite entanglement analysis and detuning dependence}

\begin{figure*}[tbh]
	\centering,
	\includegraphics[width=5.5cm]{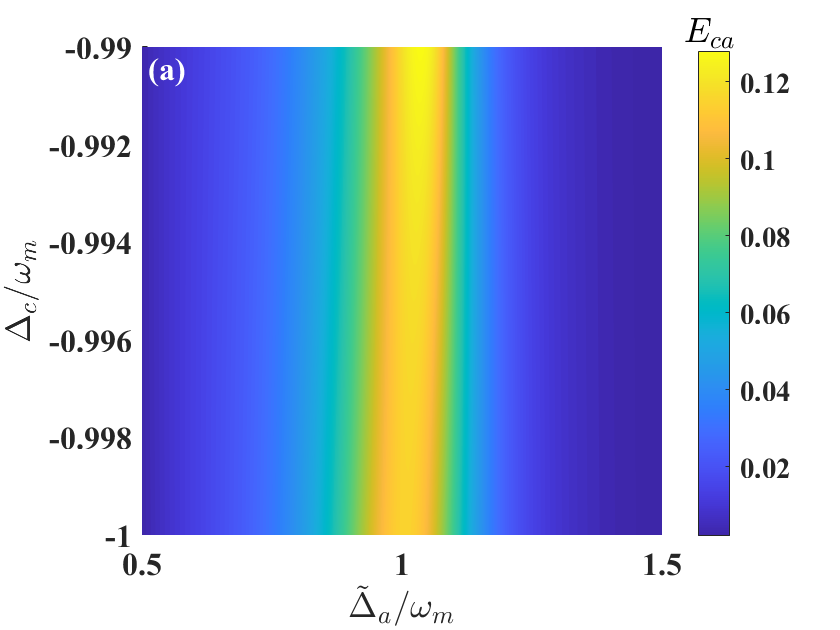}
	\includegraphics[width=5.5cm]{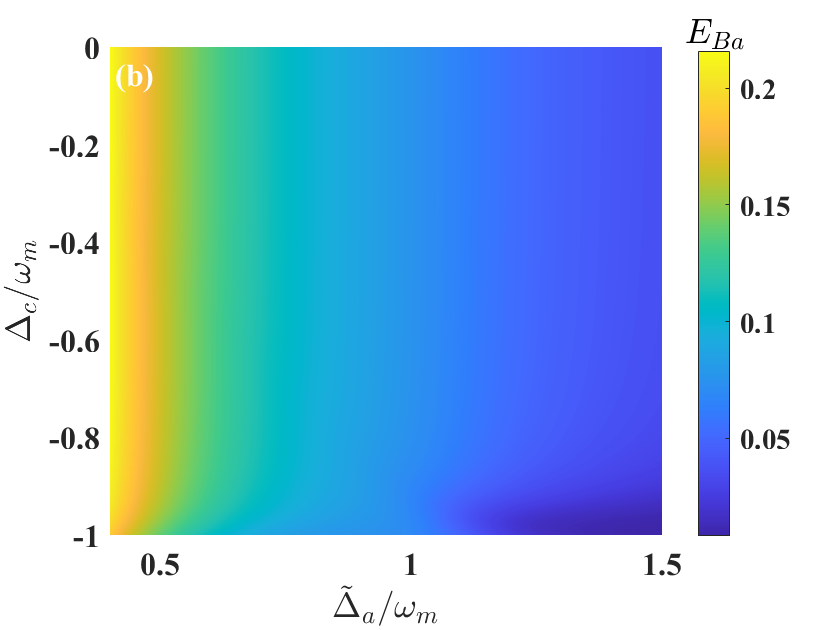}
	\includegraphics[width=5.5cm]{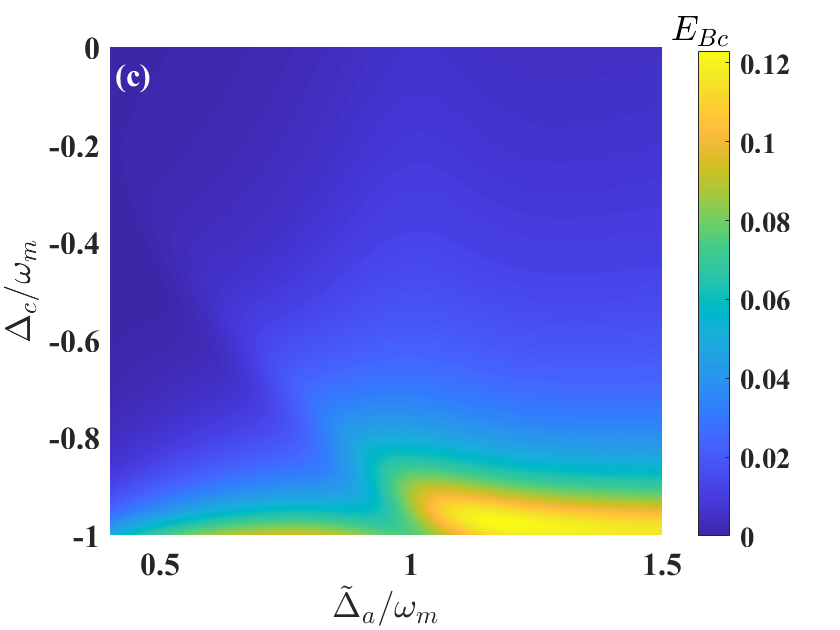}
	\includegraphics[width=5.5cm]{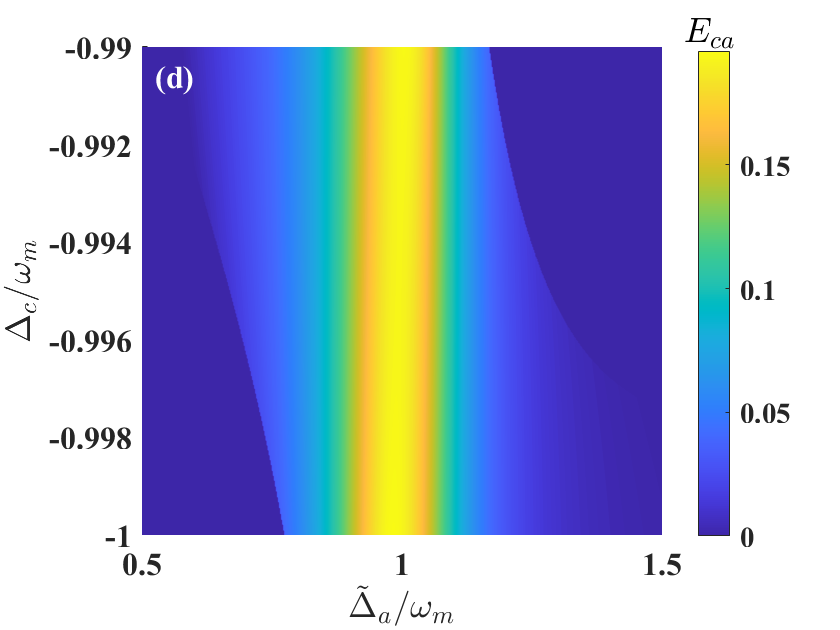}
	\includegraphics[width=5.5cm]{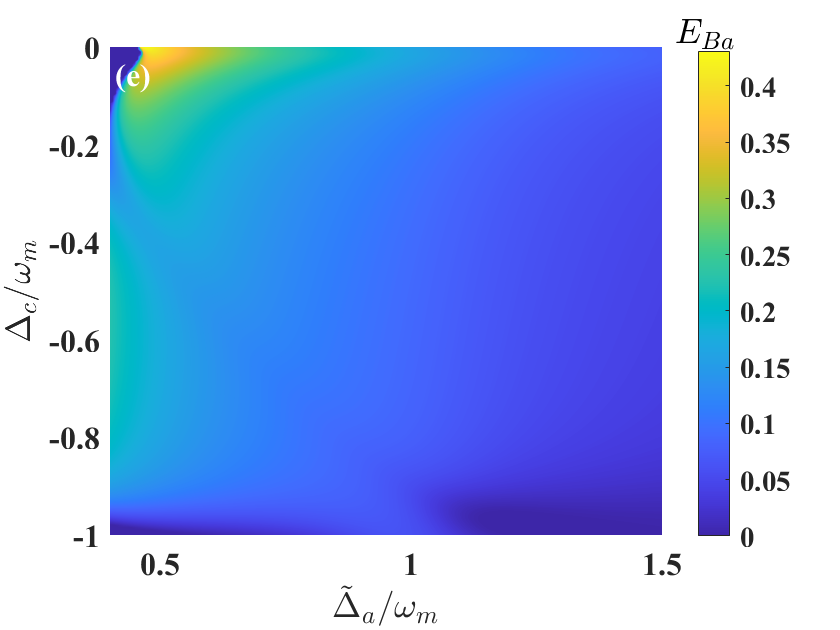}
	\includegraphics[width=5.5cm]{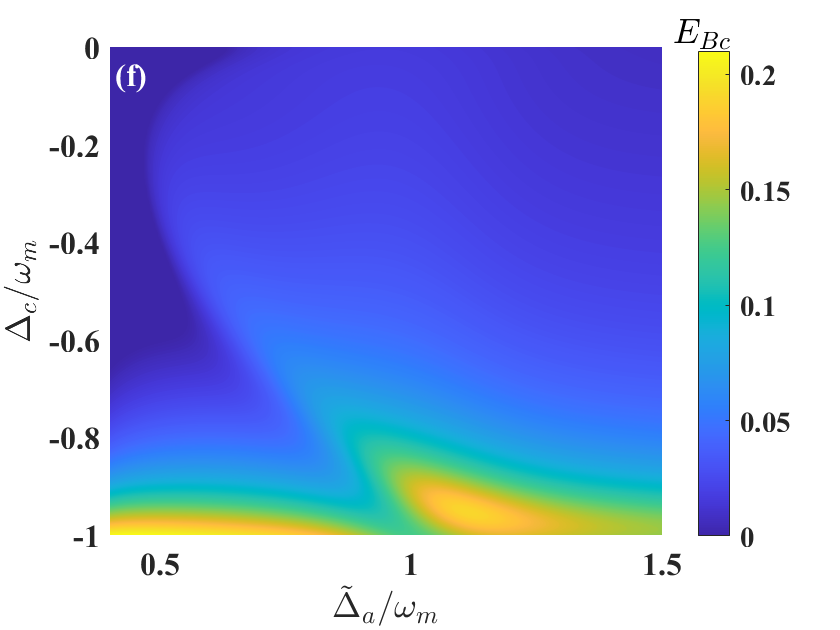}
	\caption{(a) Density plot of bipartite entanglement for photon-photon modes ($E_{ca}$) with $J/\omega_m = \num{0}$, $\kappa_j/\omega_m = \num{0.003}$ and $G_j/\omega_m = \num{0.003}$. (b) Entanglement for vibration-photon modes ($E_{Ba}$) with $J/\omega_m = \num{0}$, $\kappa_j/\omega_m = \num{0.05}$ and $G_j/\omega_m = \num{0.005}$. (c) Entanglement for vibration-photon modes ($E_{Bc}$) for $J/\omega_m = \num{0}$, as a function of normalized detunings $\tilde{\Delta}_a/\omega_m$ and ${\Delta}_c/\omega_m$. (d) Density plot of bipartite entanglement for photon-photon modes ($E_{ca}$) with $J/\omega_m = \num{0.1}$, $\kappa_j/\omega_m = \num{0.003}$ and $G_j/\omega_m = \num{0.003}$. (e) Entanglement for vibration-photon modes ($E_{Ba}$) with $J/\omega_m = \num{0.1}$, $\kappa_j/\omega_m = \num{0.05}$ and $G_j/\omega_m = \num{0.005}$. (f) Entanglement for vibration-photon modes ($E_{Bc}$) for $J/\omega_m = \num{0.1}$, as a function of normalized detunings $\tilde{\Delta}_a/\omega_m$ and ${\Delta}_c/\omega_m$. Common parameters for all subplots: $\mathcal{E}/\omega_m = 16$, $\gamma_m/\omega_m = \num{0.005}$, $g_c/\omega_m = \num{3.3e-6}$, $g_a/\omega_m = \num{2.66e-6}$, $N = \num{e6}$, $\theta=\pi$, and $T = \SI{210}{\kelvin}$.}
	\label{fig:fig5}
\end{figure*}

Here we investigate the steady-state bipartite entanglements between various subsystems, namely, cavity-cavity ($E_{ca}$), cavity $a$ and the collective molecular mode ($E_{Ba}$), and cavity $c$ and the collective molecular mode ($E_{Bc}$). \Cref{fig:fig5} provides a comprehensive overview on how these entanglement measures vary with normalized detunings $\tilde{\Delta}_a/\omega_m$ and $\Delta_c/\omega_m$. By systematically exploring these relationships, we gain important insight into the intricate interplay between cavity dynamics, molecular vibrations, and the generation of quantum correlations. This analysis allows us to identify optimal detuning regimes for maximizing and sustaining entanglement. It is worth noting that the specific parameters for $\kappa_j/\omega_m$ and $G_j/\omega_m$ are adjusted for each subplot to highlight different correlation regimes, as specified in the figure caption for \Cref{fig:fig5}(a) and \Cref{fig:fig5}(b-f).
In panels (a–c) of \Cref{fig:fig5}, we present the density plots of bipartite entanglement between subsystems in the absence of coupling between the two cavities. In panels (d–f), we show the corresponding plots in the presence of coupling between the two cavities. As it can be clearly seen in \Cref{fig:fig5}, the entanglement between subsystems is enhanced when coupling between the cavities is introduced, and suppressed when the coupling is absent. When the two cavities are uncoupled, each cavity interacts with the molecular ensemble independently. The flow of quantum correlations is limited to what can be generated locally between each cavity and the molecular mode interacts with. As a result, the overall entanglement between different subsystems remains relatively weak. However, when a coupling between the cavities is introduced, it adds energy and quantum correlations to flow between them. This coupling acts as a quantum channel that links the dynamics of the two cavities. As a result, correlations can be established between the two cavity modes via their mutual interaction with the molecular ensemble. In essence, the coupling enables a collective behaviour that strengthens entanglement by connecting different parts of the system and allowing them to interact more coherently.

As shown in \Cref{fig:fig5}a and \Cref{fig:fig5}d, the bipartite entanglement between the two cavity modes, $E_{ca}$, peaks under specific resonant conditions. This maximum strength is achieved at the red-sideband resonance, i.e., $\tilde{\Delta}_a=\omega_m$. Importantly, this cavity-cavity entanglement is mediated by the molecular collective mode, an effective coupling  emerging from the system's dynamics. This mediation allows the entanglement to reach its maximum regardless of the $\Delta_c/\omega_m$ detuning. The underlying physical principle here is that at $\tilde{\Delta}_a \approx \omega_m$ (red-detuned for cavity $a$), the molecular collective mode is efficiently cooled through resolved sideband cooling. This cooling action creates favorable conditions for transferring and establishing quantum correlations. In essence, the resonance of cavity $a$ effectively stabilizes and enhances the mediated entanglement between the two optical cavities, even as the detuning of cavity $c$ varies. In contrast, we observe that the peak of the entanglement values involving the molecular mode, i.e., $E_{Ba}$ (between cavity $a$ and mode $B$) and $E_{Bc}$ (between cavity $c$ and mode $B$), occur under different detuning conditions, i.e., $E_{Ba}$ is maximal at $\tilde{\Delta}_a/\omega_m \approx 0.5$ ($\Delta_c/\omega_m \geq -1$) for $J/\omega_m=0$ and $\tilde{\Delta}_a/\omega_m \approx 0.5$ ($\Delta_c/\omega_m \geq -0.2$) for $J/\omega_m=0.1$, while $E_{Bc}$ peaks at $\tilde{\Delta}_a/\omega_m > 1$ ($\Delta_c/\omega_m \approx -1$). This suggests a fascinating trade-off: while the optimal resonance condition ($\tilde{\Delta}_a/\omega_m \approx 1$) excels at cooling the molecular mode and maximizing mediated cavity-cavity entanglement, the direct entanglement between each cavity and the molecular collective mode remains relatively weak under these specific conditions. This behavior strongly implies an entanglement swapping or mediation process, where the molecular collective mode acts as an intermediary, facilitating robust quantum correlations between the two optical cavities while being less entangled with them.
\begin{figure*}[tbh]
	\centering
	\includegraphics[width=5.5cm]{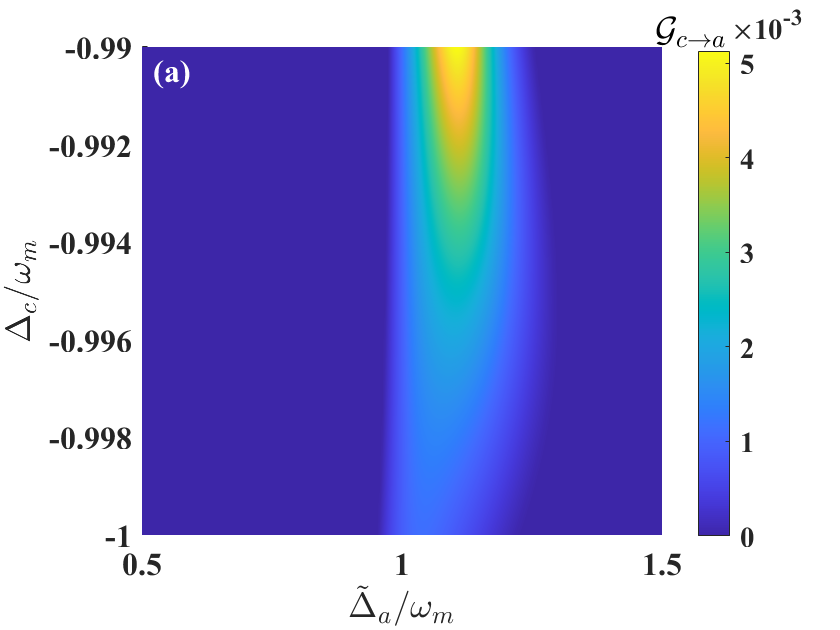}
	\includegraphics[width=5.5cm]{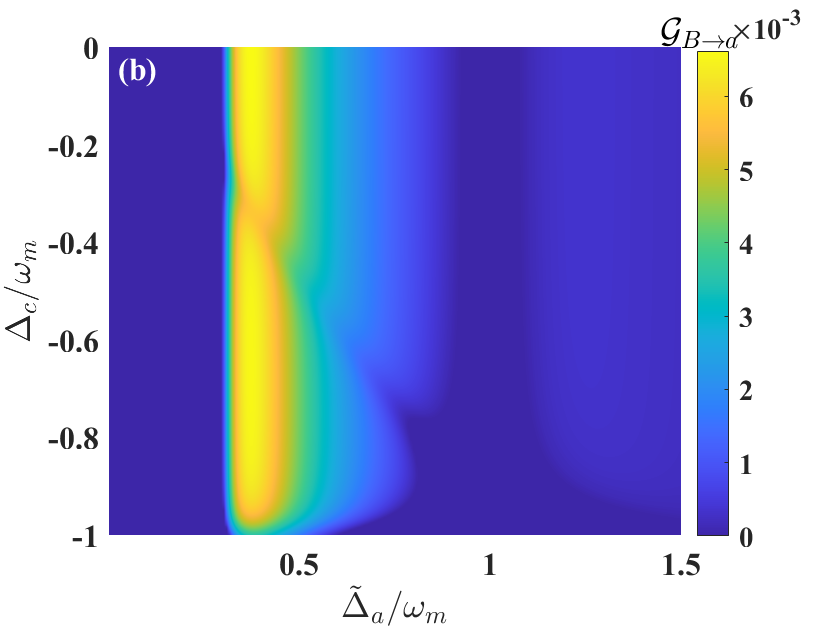}
	\includegraphics[width=5.5cm]{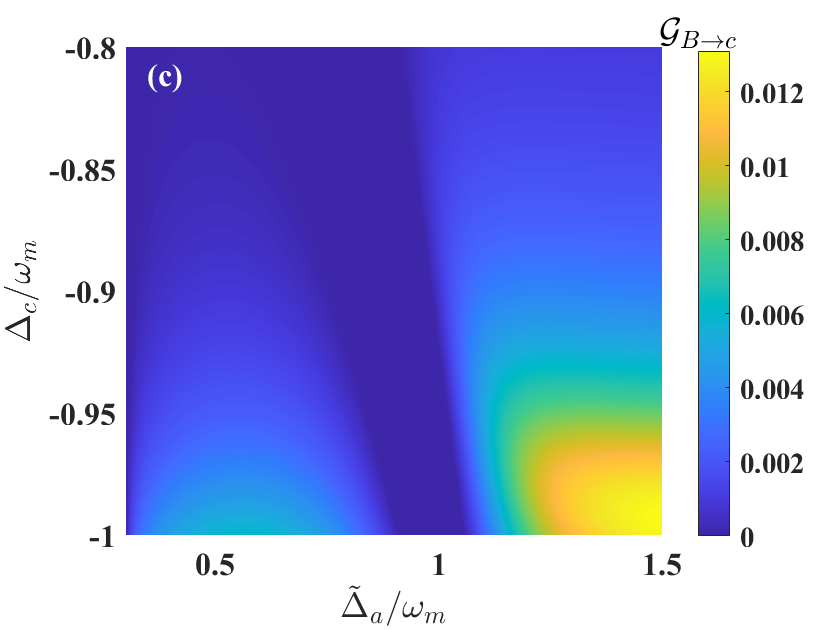}
	\includegraphics[width=5.5cm]{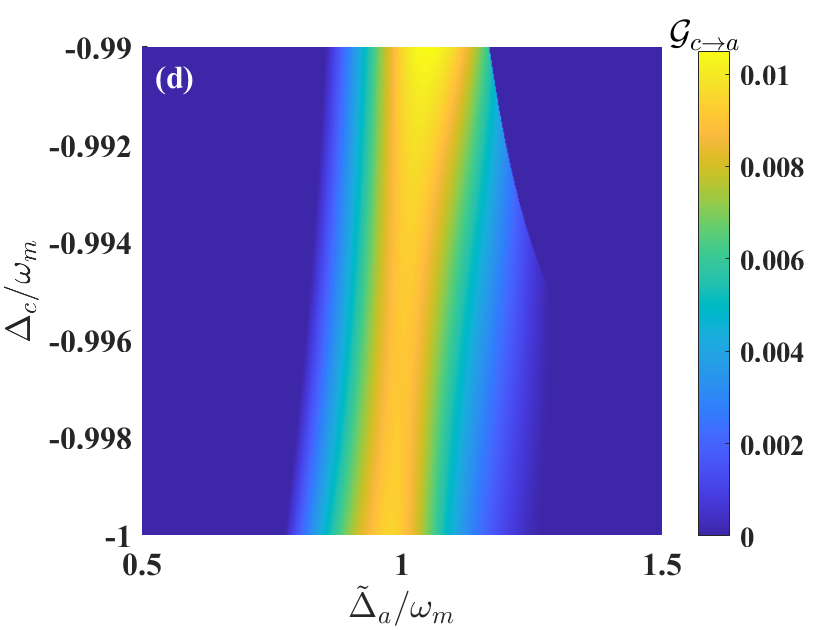}
	\includegraphics[width=5.5cm]{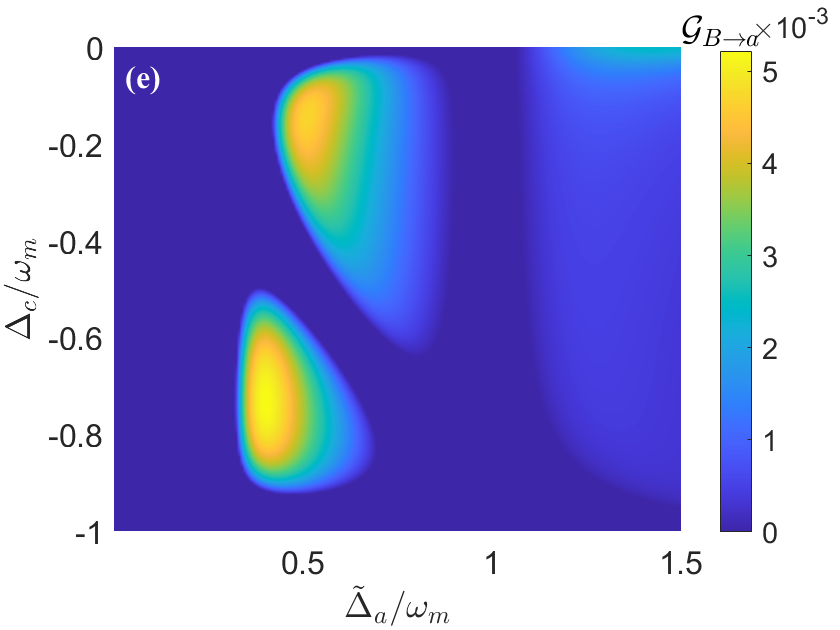}
	\includegraphics[width=5.5cm]{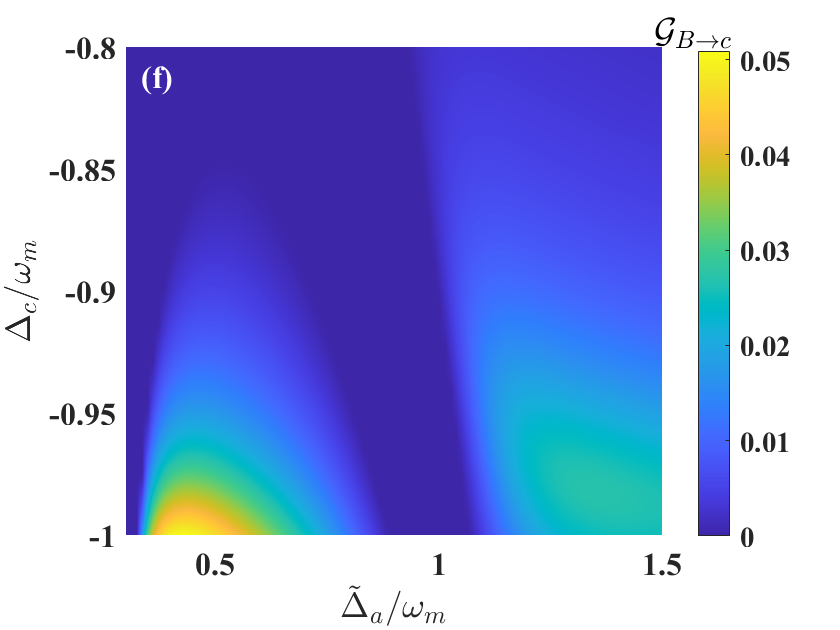}
	\caption{(a) Density plot of one-way steering $\mathcal{G}_{c\to a}$ for $J/\omega_m = \num{0}$, $\kappa_j/\omega_m = \num{0.003}$ and $G_j/\omega_m = \num{0.003}$. (b) Density plot of one-way steering $\mathcal{G}_{B\to a}$ for $J/\omega_m = \num{0}$,  $\kappa_j/\omega_m = \num{0.03}$ and $G_j/\omega_m = \num{0.003}$. (c) One-way steering $\mathcal{G}_{B\to c}$ for $J/\omega_m = \num{0}$, $\kappa_j/\omega_m = \num{0.05}$ and $G_j/\omega_m = \num{0.005}$ as a function of the normalized detunings $\tilde{\Delta}_a/\omega_m$ and ${\Delta}_c/\omega_m$. (d) Density plot of one-way steering $\mathcal{G}_{c\to a}$ for $J/\omega_m = \num{0.1}$, $\kappa_j/\omega_m = \num{0.003}$ and $G_j/\omega_m = \num{0.003}$. (e) Density plot of one-way steering $\mathcal{G}_{B\to a}$ for $J/\omega_m = \num{0.1}$, $\kappa_j/\omega_m = \num{0.03}$ and $G_j/\omega_m = \num{0.003}$. (f) One-way steering $\mathcal{G}_{B\to c}$ for $J/\omega_m = \num{0.1}$, $\kappa_j/\omega_m = \num{0.05}$ and $G_j/\omega_m = \num{0.005}$ as a function of the normalized detunings $\tilde{\Delta}_a/\omega_m$ and ${\Delta}_c/\omega_m$.   Common parameters for all subplots: $\mathcal{E}/\omega_m = 16$, $\gamma_m/\omega_m = \num{0.005}$, $g_c/\omega_m = \num{3.3e-6}$, $g_a/\omega_m = \num{2.66e-6}$, $T = \SI{210}{\kelvin}$, $\theta=\pi$, $\kappa_a= \kappa_c$, and $T = \SI{210}{\kelvin}$.}
	\label{fig:fig6}
\end{figure*}

\subsection{EPR steering dynamics and detuning dependence}
Next, we analyze the one-way EPR steering, $\mathcal{G}$, focusing on its directional properties. \Cref{fig:fig6}(a-f) illustrate the one-way steering from cavity mode $c$ to cavity mode $a$ ($\mathcal{G}_{c\to a}$), from collective molecular mode $B$ to cavity mode $a$ ($\mathcal{G}_{B\to a}$), and from mode $B$ to cavity mode $c$ ($\mathcal{G}_{B\to c}$), all plotted against normalized detunings $\tilde{\Delta}_a/\omega_m$ and $\Delta_c/\omega_m$. A clear optimal region for $\mathcal{G}_{c\to a}$ emerges in \Cref{fig:fig6}(a) and \Cref{fig:fig6}(d) when $\tilde{\Delta}_a/\omega_m \approx 1$ and $\Delta_c/\omega_m > -1$ for $J/\omega_m=0$ and $\Delta_c/\omega_m \geq -1$ for $J/\omega_m=0.1$. This observation underscores that maximum entanglement (as discussed previously) and steering between cavity modes are achieved when cavity decay rates ($\kappa_j/\omega_m$) and coupling strengths ($G_j/\omega_m$) are significantly lower than the mechanical damping rate ($\gamma_m/\omega_m$). Specifically, for the plotted scenarios, $\kappa_j/\omega_m = \num{0.003}$ and $G_j/\omega_m = \num{0.003}$, with $\gamma_m/\omega_m = \num{0.005}$. A reduced cavity decay rate effectively preserves quantum correlations by minimizing decoherence, thereby allowing one cavity mode to steer the conditional state of the other. Interestingly, one-way EPR steering from the molecular mode to cavity $a$ (denoted as $\mathcal{G}_{B \to a}$) reaches its peak at $\tilde{\Delta}_a/\omega_m \approx 0.5$ (with $\Delta_c/\omega_m > -1$) for $J/\omega_m = 0$, and within the range $-0.9 \leq \Delta_c/\omega_m \leq -0.7$ (or $-0.4 \leq \Delta_c/\omega_m \leq -0.1$) for $J/\omega_m = 0.1$. In contrast, $\mathcal{G}_{B \to c}$ is maximized at $\tilde{\Delta}_a/\omega_m > 1$ (with $\Delta_c/\omega_m \approx -1$) for $J/\omega_m = 0$, and at $\tilde{\Delta}_a/\omega_m < 1$ (also with $\Delta_c/\omega_m \approx -1$) for $J/\omega_m = 0.1$. This highlights distinct optimal conditions for different steering directions, reflecting the asymmetric responses of the system.
\begin{figure*}[tbh]
	\centering
	\includegraphics[width=5.5cm]{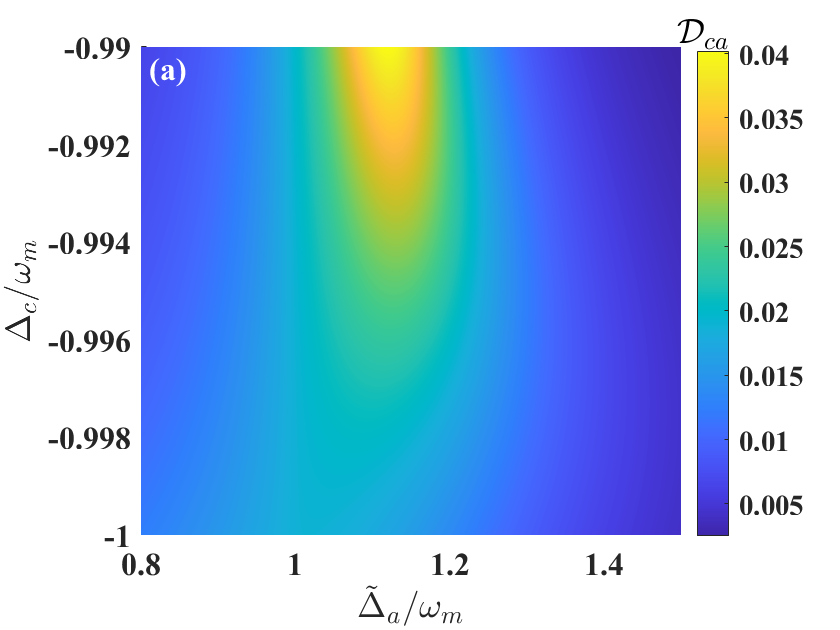}
	\includegraphics[width=5.5cm]{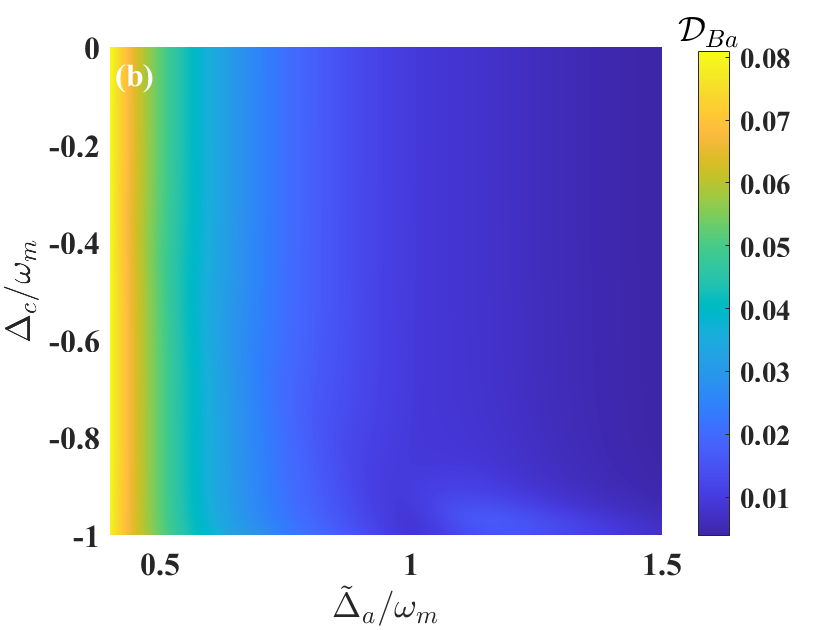}
	\includegraphics[width=5.5cm]{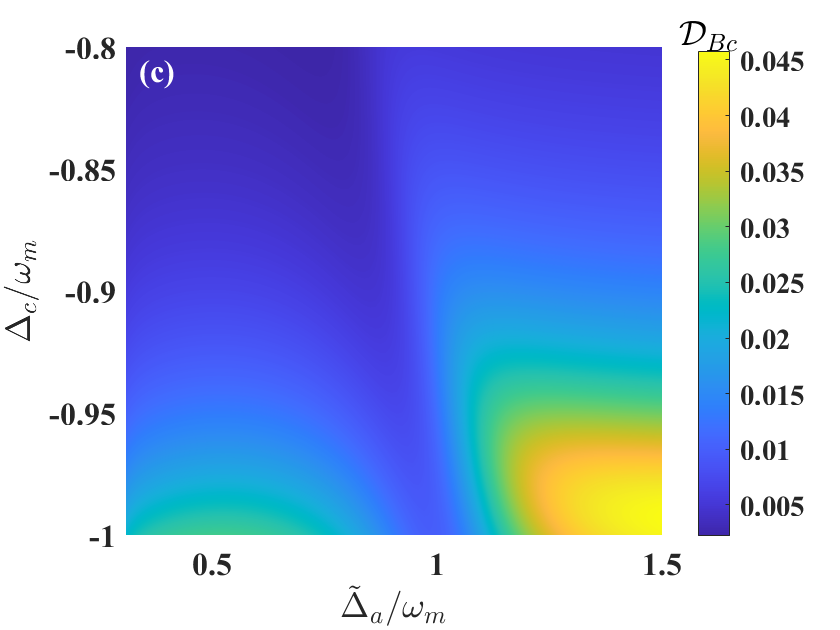}
	\includegraphics[width=5.5cm]{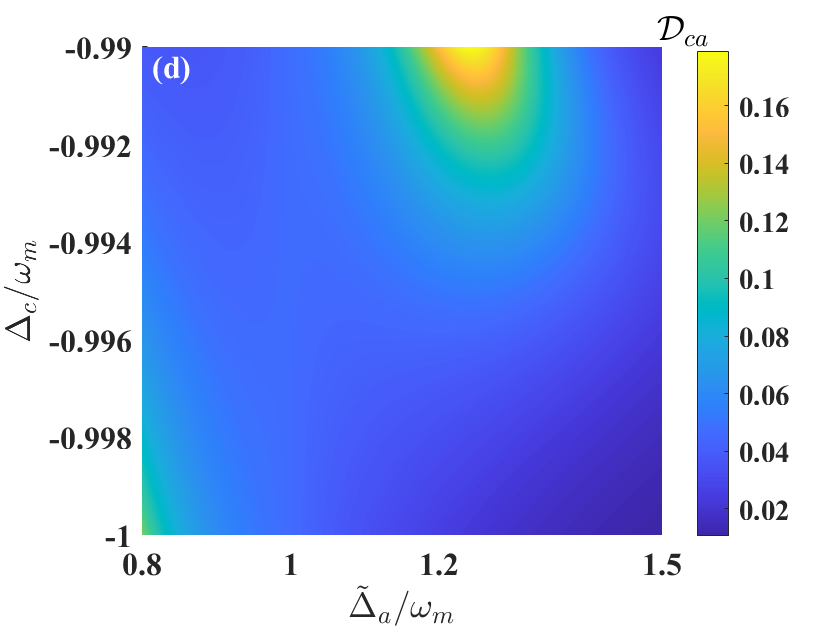}
	\includegraphics[width=5.5cm]{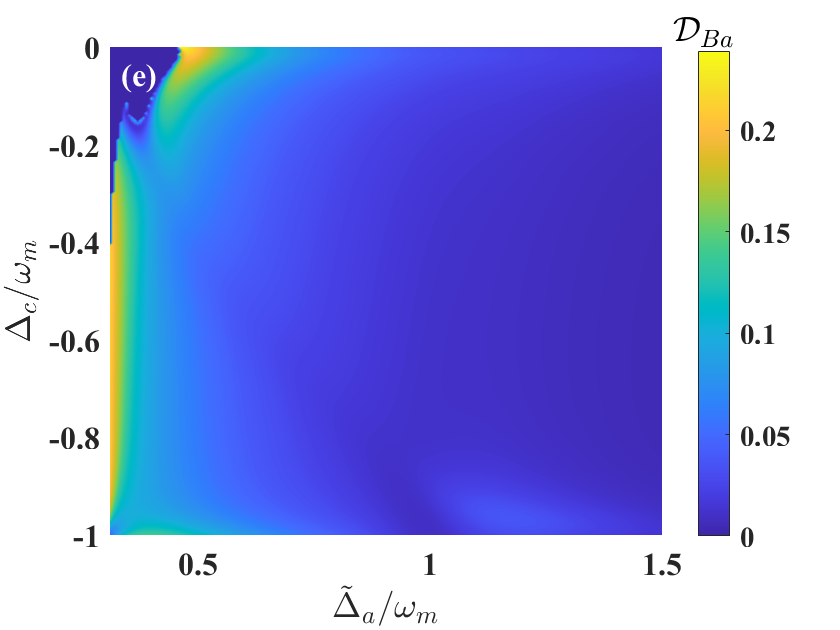}
	\includegraphics[width=5.5cm]{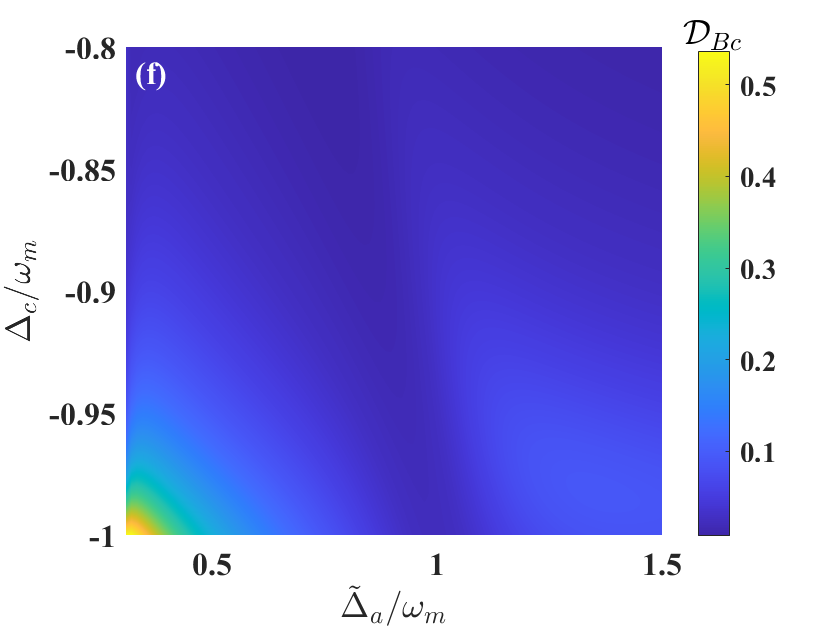}
	\caption{(a) Density plot of quantum discord $\mathcal{D}_{ca}$ between cavity modes $a$ and $c$, for $J/\omega_m = \num{0}$, $\kappa_j/\omega_m = \num{0.005}$ and $G_j/\omega_m = \num{0.003}$. This is the range that is giving something similar to the figure when J=0. (b) Quantum discord $\mathcal{D}_{Ba}$ between the molecular collective mode $B$ and cavity mode $a$, for $J/\omega_m = \num{0}$, $\kappa_j/\omega_m = \num{0.05}$ and $G_j/\omega_m = \num{0.005}$. (c) Quantum discord $\mathcal{D}_{Bc}$ between the molecular collective mode $B$ and cavity mode $c$, for $J/\omega_m = \num{0}$ as a function of normalized detunings $\tilde{\Delta}_a/\omega_m$ and ${\Delta}_c/\omega_m$. (d) Density plot of quantum discord $\mathcal{D}_{ca}$ between cavity modes $a$ and $c$, for $J/\omega_m = \num{0.1}$, $\kappa_j/\omega_m = \num{0.005}$ and $G_j/\omega_m = \num{0.003}$. This is the range that is giving something similar to the figure when J=0. (e) Quantum discord $\mathcal{D}_{Ba}$ between the molecular collective mode $B$ and cavity mode $a$, for $J/\omega_m = \num{0.1}$, $\kappa_j/\omega_m = \num{0.05}$ and $G_j/\omega_m = \num{0.005}$. (f) Quantum discord $\mathcal{D}_{Bc}$ between the molecular collective mode $B$ and cavity mode $c$, for $J/\omega_m = \num{0.1}$ as a function of normalized detunings $\tilde{\Delta}_a/\omega_m$ and ${\Delta}_c/\omega_m$. Common parameters for all subplots: $\mathcal{E}/\omega_m = 16$, $\gamma_m/\omega_m = \num{0.005}$, $g_c/\omega_m = \num{3.3e-6}$, $g_a/\omega_m = \num{2.66e-6}$, $T = \SI{210}{\kelvin}$, and $\theta=\pi$.}
	\label{fig:fig7}
\end{figure*}

As mentioned earlier, when the coupling between the two cavities is introduced, it allows quantum correlations to flow more freely between them. This increased interaction not only strengthens the overall entanglement shared by the subsystems but also improves EPR steering, which depends on the ability of one subsystem to influence (or “steer”) the state of another. Essentially, the cavity coupling creates additional pathways for quantum information to be exchanged, enhancing both entanglement and steering simultaneously. Although the introduction of inter-cavity coupling $J$ enhances the overall entanglement in the system, it does not necessarily lead to enhanced EPR steering between cavity mode $a$ and the molecular mode $B$. This behaviour can be understood from the bilinear interaction between the involving modes as seen in the  Hamiltonian \Cref{eq:3}. This weak bilinear interaction between cavity mode $a$ and the molecular vibrational mode $B$ is of the dispersive radiation-pressure type, described by the nonlinear term $G_aa^\dagger a(B^\dagger + B)$, whereas cavity mode $c$ couples to $B$ through a linear beam-splitter-like interaction, $G_c(c^\dagger + c)(B^\dagger + B)$. These two forms of couplings have different efficiencies in generating quantum correlations. When $J \neq 0$, the cavity modes $a$ and $c$ become hybridized due to the inter-cavity tunneling term $J(a^\dagger c + a c^\dagger)$. As a result, the photonic excitations in mode $a$ become delocalized over both cavities. This hybridization redistributes the optomechanical interaction, effectively weakening the direct coupling between $a$ and $B$. 
\begin{figure*}[tbh]
	\centering
	\includegraphics[width=5.5cm]{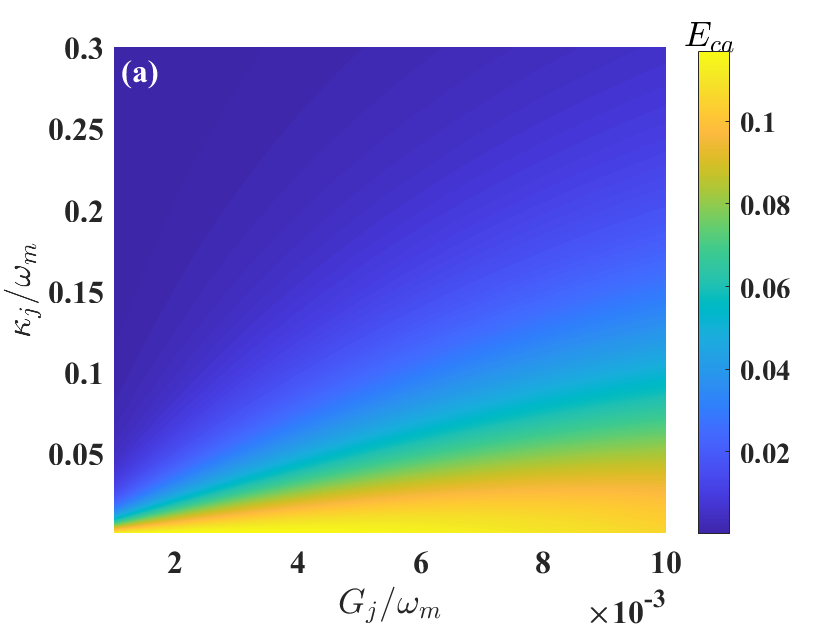}
	\includegraphics[width=5.5cm]{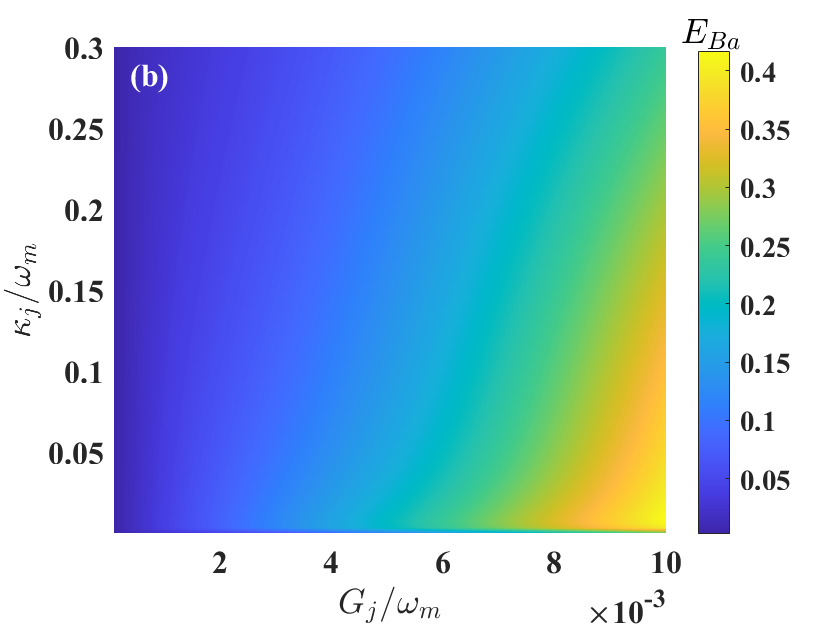}
	\includegraphics[width=5.5cm]{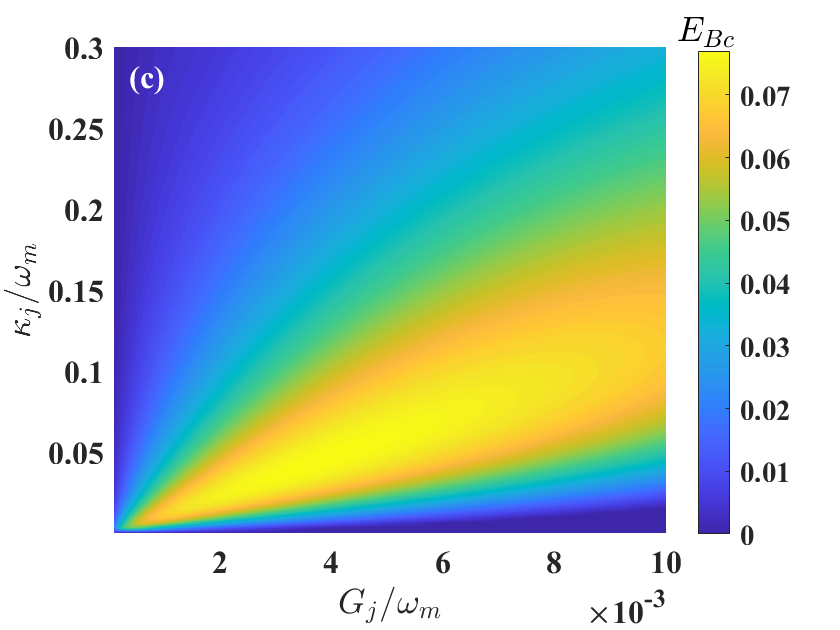}
	\includegraphics[width=5.5cm]{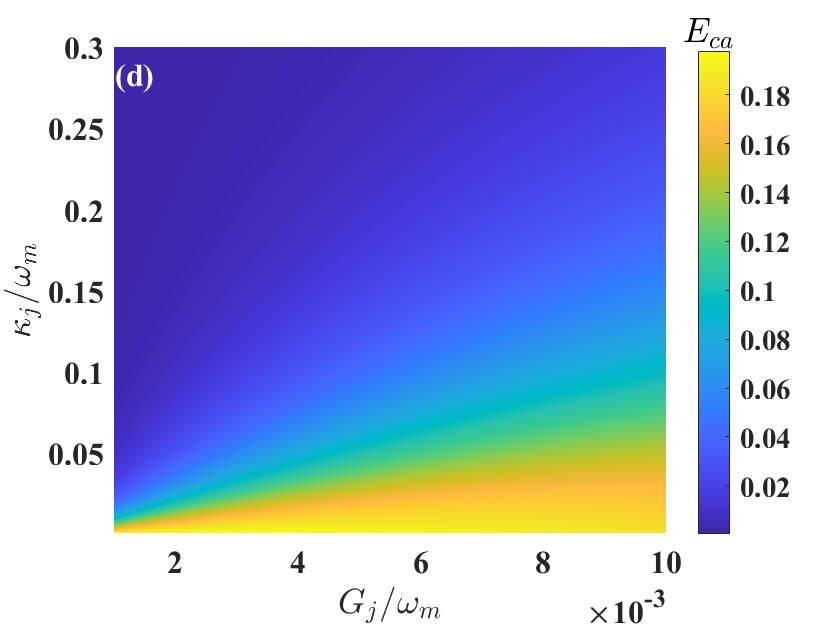}
	\includegraphics[width=5.5cm]{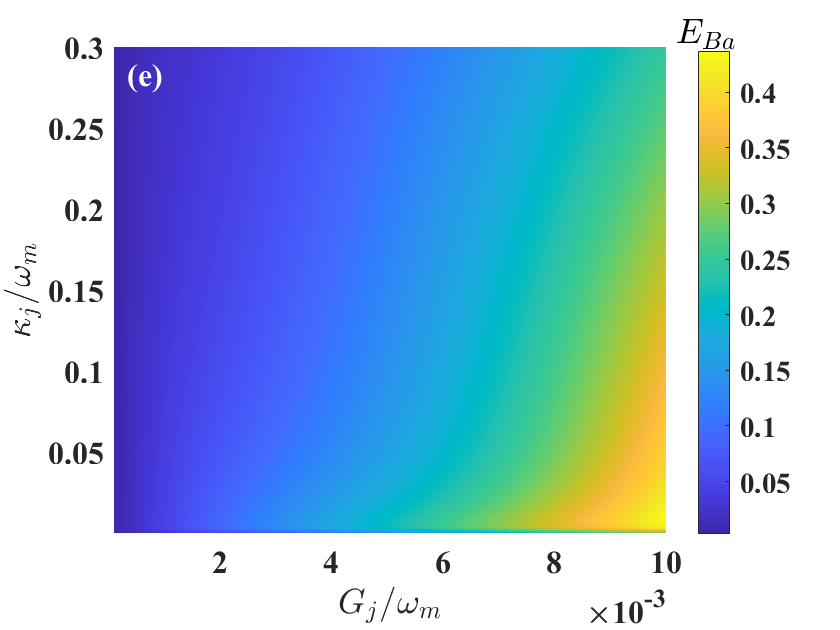}
	\includegraphics[width=5.5cm]{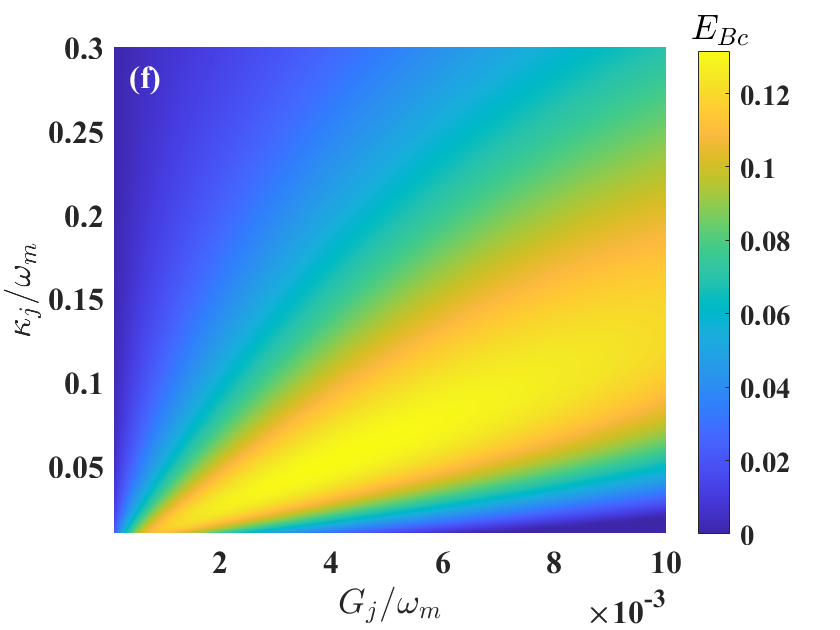}
	\caption{(a) Density plot of bipartite entanglement $E_{ca}$ between cavity modes $a$ and $c$, for $J/\omega_m = \num{0}$, $\tilde{\Delta}_a/\omega_m = 1$ and ${\Delta}_c/\omega_m = -1$. (b) Entanglement $E_{Ba}$ between the molecular collective mode $B$ and cavity mode $a$, for $J/\omega_m = \num{0}$,  $\tilde{\Delta}_a/\omega_m = \num{0.5}$ and ${\Delta}_c/\omega_m = -0.8$. (c) Entanglement $E_{Bc}$ between the molecular collective mode $B$ and cavity mode $c$, for $J/\omega_m = \num{0}$, $\tilde{\Delta}_a/\omega_m = 1$ and ${\Delta}_c/\omega_m = -1$. (d) Density plot of bipartite entanglement $E_{ca}$ between cavity modes $a$ and $c$, for $J/\omega_m = \num{0.1}$,  $\tilde{\Delta}_a/\omega_m = 1$ and ${\Delta}_c/\omega_m = -1$. (e) Entanglement $E_{Ba}$ between the molecular collective mode $B$ and cavity mode $a$, for $J/\omega_m = \num{0.1}$,  $\tilde{\Delta}_a/\omega_m = \num{0.5}$ and ${\Delta}_c/\omega_m = -0.8$. (f) Entanglement $E_{Bc}$ between the molecular collective mode $B$ and cavity mode $c$, for $J/\omega_m = \num{0.1}$, $\tilde{\Delta}_a/\omega_m = 1$ and ${\Delta}_c/\omega_m = -1$. Common parameters for all subplots: $\mathcal{E}/\omega_m = 16$, $\gamma_m/\omega_m = \num{0.005}$, $g_c/\omega_m = \num{3.3e-6}$, $g_a/\omega_m = \num{2.66e-6}$, $T = \SI{210}{\kelvin}$, and $\theta=\pi$.}
	\label{fig:fig8}
\end{figure*}

Moreover, EPR steering is a directional form of quantum correlation and is generally more sensitive to asymmetries in system parameters and interaction types. While entanglement is a symmetric quantity, EPR steering from $B \rightarrow a$ (or vice versa) relies on how well measurements on one subsystem can infer conditional states of the other. The nonlinear nature of the $a$--$B$ interaction, combined with mode delocalization caused by nonzero $J$, diminishes this directional predictability. In contrast, the cavity mode $c$ maintains a direct and linear interaction with $B$, and benefits from the additional coupling to mode $a$ via $J$, which can enhance the steering in channels involving $c$. Therefore, the asymmetry in interaction types and the redistribution of coupling strength explain why the EPR steering between cavity mode $a$ and the molecular mode $B$ is not enhanced when the cavity-cavity coupling is introduced.
\subsection{Gaussian Quantum Discord behavior and detuning dependence}
Our exploration extends to Gaussian Quantum Discord (GQD), a robust measure that captures both entangled and separable quantum correlations. \Cref{fig:fig7} presents density plots for the quantum discord between the cavity modes ($\mathcal{D}_{ca}$), between the molecular collective mode and the cavity $a$ ($\mathcal{D}_{Ba}$), and between the molecular collective mode and cavity $c$ ($\mathcal{D}_{Bc}$), as a function of normalized detunings. The parameters for subplots (a) and (b-f) are distinct, as specified in the figure caption. The quantum discord $\mathcal{D}_{ca}$ between the two cavity modes attains its maximum in the off-resonant regime, specifically when both cavity detunings satisfy $\tilde{\Delta}_a/\omega_m > 1$ and $\Delta_c/\omega_m > -1$ (see \Cref{fig:fig7}(a) and (d)). This intriguing observation suggests that quantum correlations, as measured by discord, can robustly persist and even be enhanced when both cavities are far-detuned from the molecular mediator (\Cref{fig:fig7}(a)), a valuable feature for quantum information tasks. Furthermore, \Cref{fig:fig7}(b) shows that the optimal quantum discord between cavity $a$ and the molecular collective mode occurs at $\tilde{\Delta}_a/\omega_m \approx 0.35$, where their interaction is neither fully resonant nor completely off-resonant. Intriguingly, this maximum in discord exhibits remarkable robustness across a wide range of detunings for cavity $c$, underscoring the inherently local nature of quantum discord and its resilience to environmental perturbations. \Cref{fig:fig7}(c) shows that $\mathcal{D}_{Bc}$ reaches its maximum when cavity $a$ is off-resonant with the collective molecular mode, specifically for $\tilde{\Delta}_a/\omega_m > 1$ and $\Delta_c/\omega_m \in [-1, -0.9]$ when $J/\omega_m = 0$, and for $\tilde{\Delta}_a/\omega_m < 0.5$ and $\Delta_c/\omega_m \in [-1, -0.95]$ when $J/\omega_m = 0.1$. In this regime, the molecular collective mode interacts weakly with the cavity $a$ when $J/\omega_m = 0.1$, which allows the correlations between the molecular collective mode and the cavity $c$ to be maximally enhanced (see \Cref{fig:fig7}(f)).

The enhancement of quantum discord between cavity mode $a$ and the molecular mode $B$ in the presence of inter-cavity coupling ($J \neq 0$) can be attributed to the redistribution of quantum correlations across the hybrid system. Although the direct interaction between $a$ and $B$ is of the dispersive radiation-pressure type, introducing a finite coupling $J$ between cavities $a$ and $c$ allows additional correlation pathways through mode $c$, which is linearly coupled to $B$. This hybridization enables indirect correlations to be established between $a$ and $B$, thereby increasing quantum discord. Unlike entanglement and EPR steering, which are more sensitive to the symmetry and type of interactions, quantum discord captures a broader class of quantum correlations, including those present in separable (non-entangled) states. Therefore, even in regimes where EPR steering between $a$ and $B$ is not enhanced, quantum discord can still increase due to the presence of non-classical correlations mediated by the coupled cavity structure.
This behavior highlights the distinct nature of quantum discord compared to entanglement and steering: while entanglement quantifies symmetric, shared non-classicality, and steering captures directional control between subsystems, quantum discord reflects more general quantum correlations that can persist under asymmetric and indirect interactions.
\begin{figure*}[tbh]
	\centering
	\includegraphics[width=5.5cm]{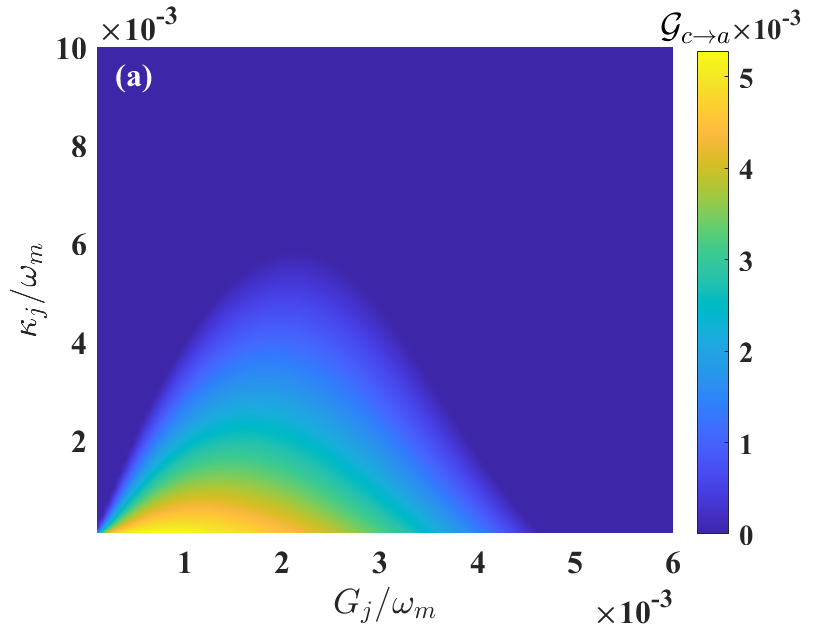}
	\includegraphics[width=5.5cm]{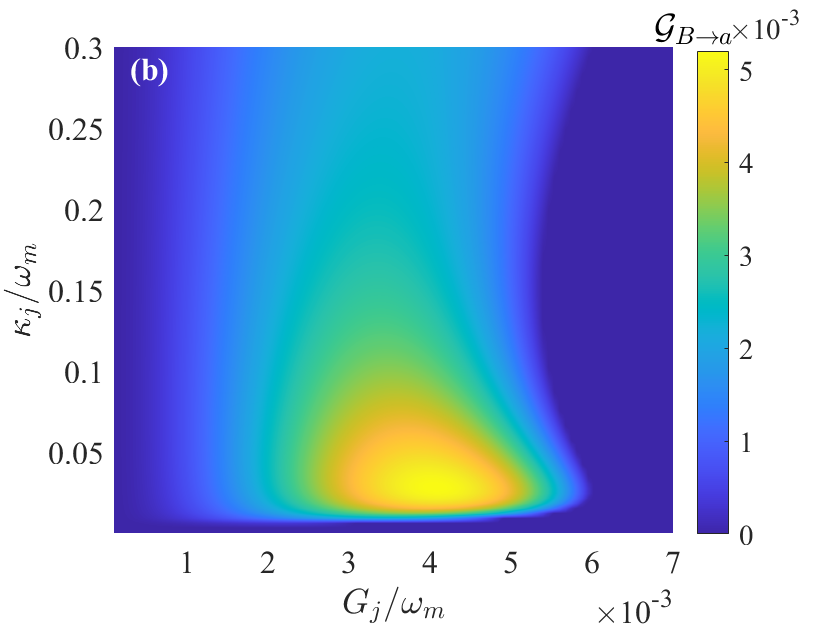}
	\includegraphics[width=5.5cm]{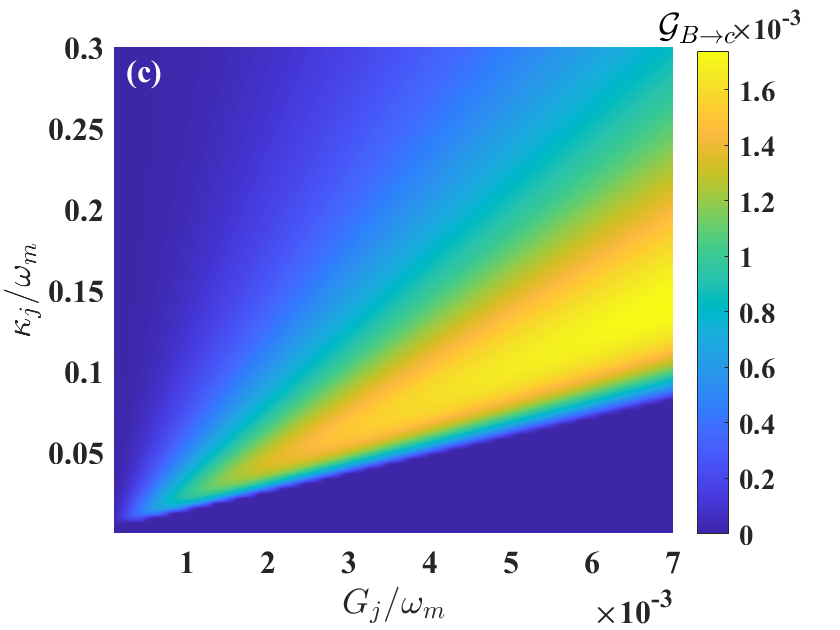}
	\includegraphics[width=5.5cm]{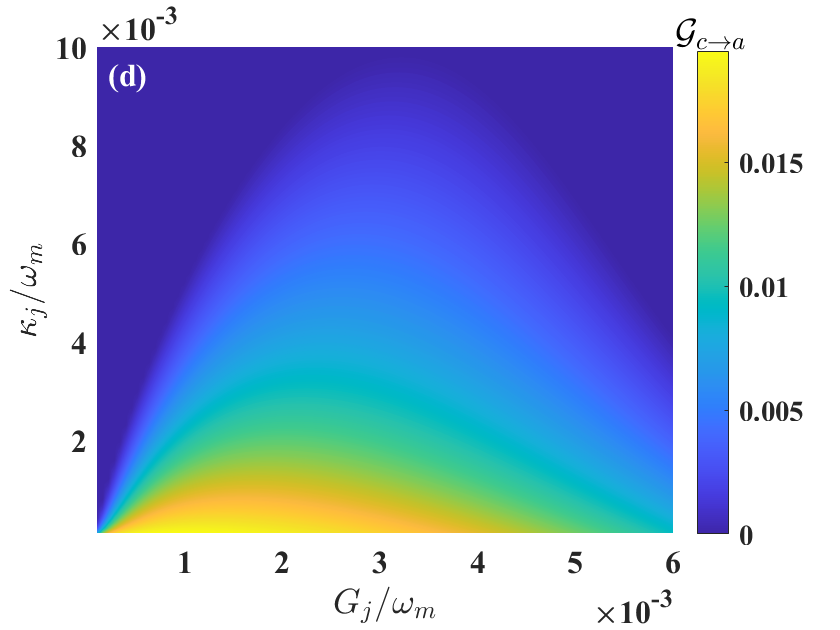}
	\includegraphics[width=5.5cm]{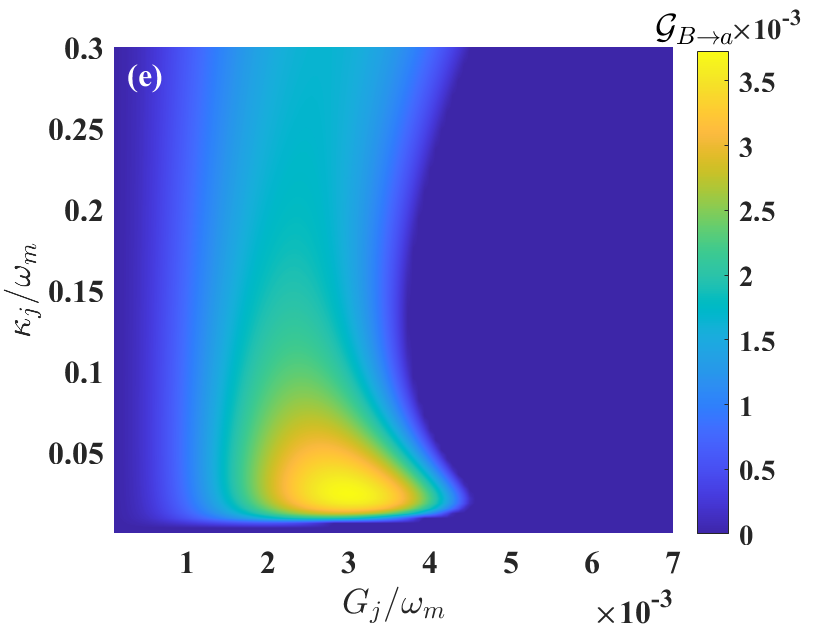}
	\includegraphics[width=5.5cm]{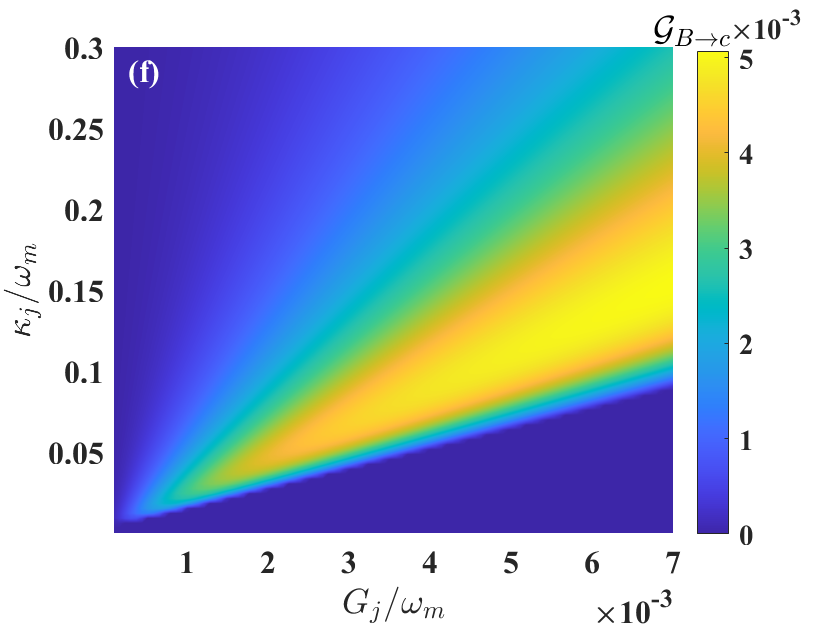}
	\caption{(a) Density plot of steering $\mathcal{G}_{c\to a}$ from cavity $c$ to $a$, for $J/\omega_m = \num{0}$,  $\tilde{\Delta}_a/\omega_m = 1$ and ${\Delta}_c/\omega_m = -1$. (b) Density plot of steering $\mathcal{G}_{B\to a}$ from molecular mode $B$ to cavity $a$, for $J/\omega_m = \num{0}$,  $\tilde{\Delta}_a/\omega_m = \num{0.5}$ and ${\Delta}_c/\omega_m = -0.8$. $\mathcal{G}_{B\to a}$ persists robustly even at higher cavity decay rates. (c) Density plot of steering $\mathcal{G}_{B\to c}$ from molecular mode $B$ to cavity $c$, for $J/\omega_m = \num{0}$, $\tilde{\Delta}_a/\omega_m = 1$ and ${\Delta}_c/\omega_m = -1$.(d) Density plot of steering $\mathcal{G}_{c\to a}$ from cavity $c$ to $a$, for  $J/\omega_m = \num{0.1}$, $\tilde{\Delta}_a/\omega_m = 1$ and ${\Delta}_c/\omega_m = -1$. (e) Density plot of steering $\mathcal{G}_{B\to a}$ from molecular mode $B$ to cavity $a$, for  $J/\omega_m = \num{0.1}$,  $\tilde{\Delta}_a/\omega_m = \num{0.5}$ and ${\Delta}_c/\omega_m = -0.8$. $\mathcal{G}_{B\to a}$ persists robustly even at higher cavity decay rates. (f) Density plot of steering $\mathcal{G}_{B\to c}$ from molecular mode $B$ to cavity $c$, for  $J/\omega_m = \num{0.1}$, $\tilde{\Delta}_a/\omega_m = 1$ and ${\Delta}_c/\omega_m = -1$. Common parameters for all subplots: $\mathcal{E}/\omega_m = 16$, $\gamma_m/\omega_m = \num{0.005}$, $g_c/\omega_m = \num{3.3e-6}$, $g_a/\omega_m = \num{2.66e-6}$, $N = \num{e6}$, $T = \SI{210}{\kelvin}$, and $\theta=\pi$.}
	\label{fig:fig9}
\end{figure*}
\subsection{Impact of system parameters on quantum correlations}
To further have an insight into the system performance, \Cref{fig:fig8} illustrates the bipartite entanglement values ($E_{ca}$, $E_{Ba}$, and $E_{Bc}$) as functions of the cavity decay rate $\kappa_j/\omega_m$ and the collective optomechanical coupling strength $G_j/\omega_m$. As anticipated, entanglement is enhanced with increasing collective optomechanical coupling strength $G_j$ and, conversely, decreases with increasing cavity decay rate $\kappa_j$ (\Cref{fig:fig8}(a)). This direct relationship stems from the proportionality between the effective induced coupling  and the collective optomechanical coupling strengths $G_j$. A particularly promising feature, visible in \Cref{fig:fig8}(b-c) and \Cref{fig:fig8}(e-f), is that the molecular collective mode--cavity mode entanglement ($E_{Ba}$) can persist even at relatively high cavity decay rates. This feature is highly advantageous for experimental realizations as it relaxes the stringent requirements for high-quality factor cavities, a common hurdle in optomechanics.

\begin{figure*}[tbh]
	\centering
	\includegraphics[width=5.5cm]{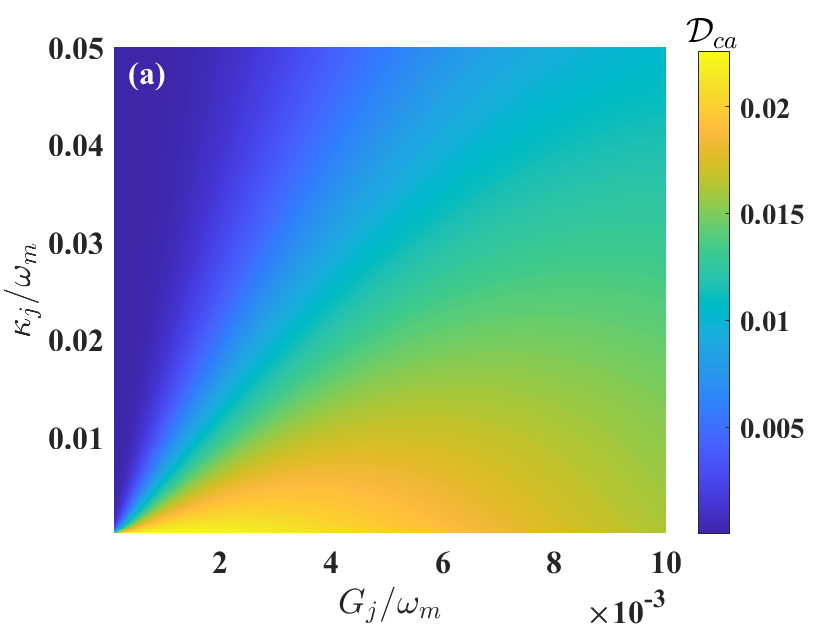}
	\includegraphics[width=5.5cm]{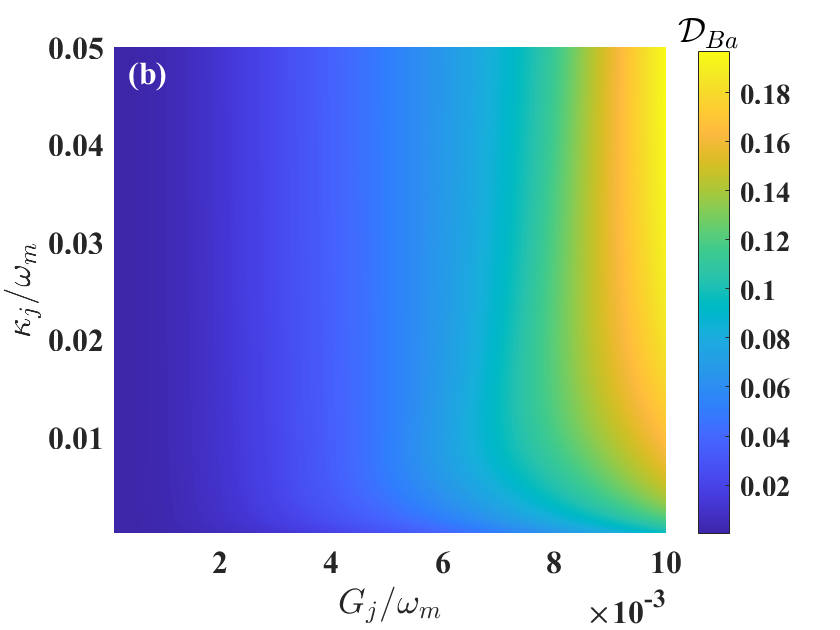}
	\includegraphics[width=5.5cm]{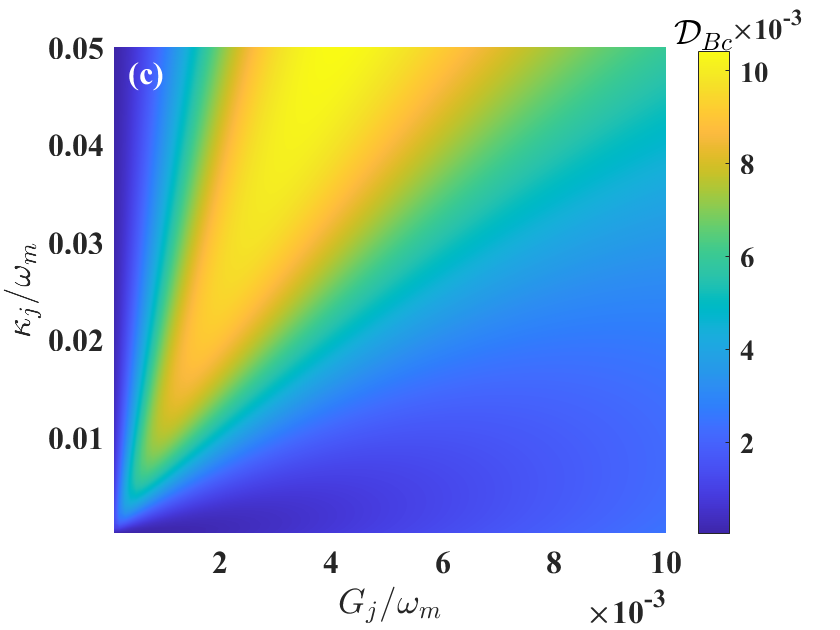}
	\includegraphics[width=5.5cm]{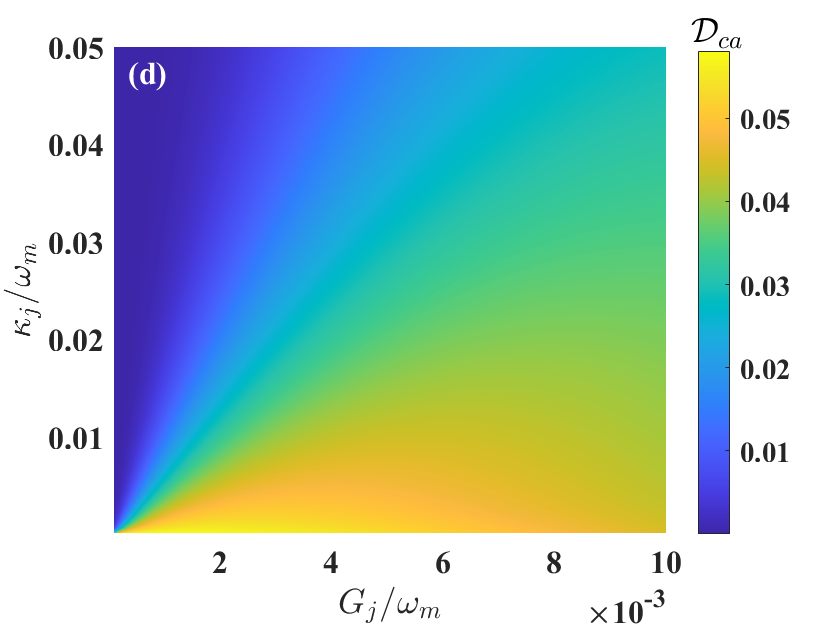}
	\includegraphics[width=5.5cm]{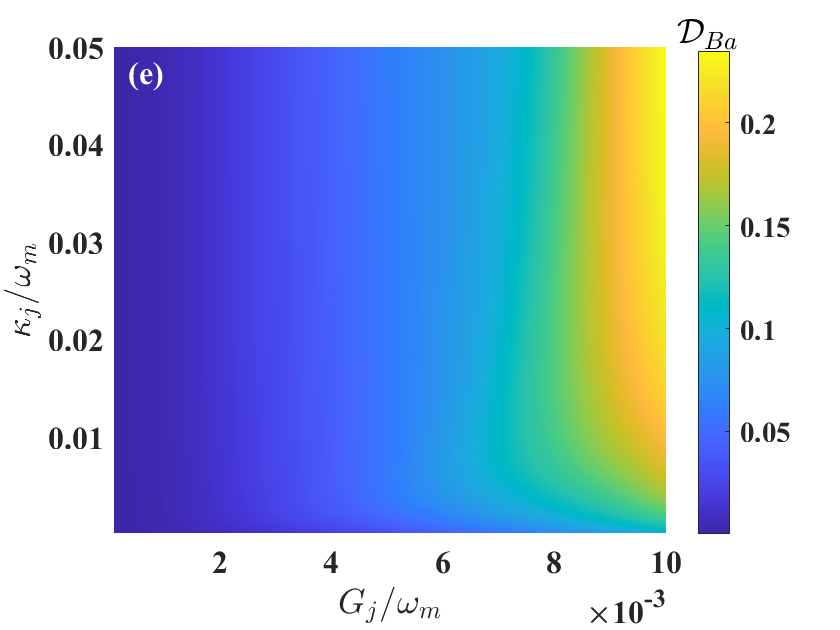}
	\includegraphics[width=5.5cm]{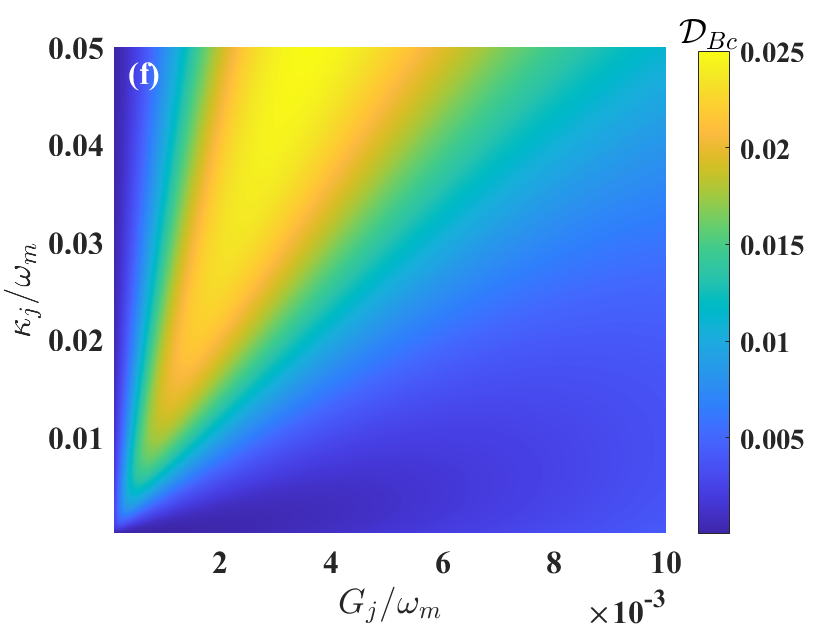}
	\caption{(a) Density plot of quantum discord $\mathcal{D}_{ca}$ between cavity modes $a$ and $c$, for $J/\omega_m = \num{0}$, $\tilde{\Delta}_a/\omega_m = 1$ and ${\Delta}_c/\omega_m = -1$. (b) Quantum discord $\mathcal{D}_{Ba}$ between the molecular collective mode $B$ and cavity mode $a$, for $J/\omega_m = \num{0}$, $\tilde{\Delta}_a/\omega_m = \num{0.5}$ and ${\Delta}_c/\omega_m = -0.8$. (c) Quantum discord $\mathcal{D}_{Bc}$ between the molecular collective mode $B$ and cavity mode $c$, for $J/\omega_m = \num{0}$, $\tilde{\Delta}_a/\omega_m = 1$ and ${\Delta}_c/\omega_m = -1$. (d) Density plot of quantum discord $\mathcal{D}_{ca}$ between cavity modes $a$ and $c$, for $J/\omega_m = \num{0.1}$, $\tilde{\Delta}_a/\omega_m = 1$ and ${\Delta}_c/\omega_m = -1$. (e) Quantum discord $\mathcal{D}_{Ba}$ between the molecular collective mode $B$ and cavity mode $a$, for $J/\omega_m = \num{0.1}$, $\tilde{\Delta}_a/\omega_m = \num{0.5}$ and ${\Delta}_c/\omega_m = -0.8$. (f) Quantum discord $\mathcal{D}_{Bc}$ between the molecular collective mode $B$ and cavity mode $c$, for $J/\omega_m = \num{0.1}$, $\tilde{\Delta}_a/\omega_m = 1$ and ${\Delta}_c/\omega_m = -1$. Common parameters for all subplots: $\mathcal{E}/\omega_m = 16$, $\gamma_m/\omega_m = \num{0.005}$, $g_c/\omega_m = \num{3.3e-6}$, $g_a/\omega_m = \num{2.66e-6}$, $T = \SI{210}{\kelvin}$, and $\theta=\pi$.}
	\label{fig:fig10}
\end{figure*}

\Cref{fig:fig9}(a--f) further investigate the one-way EPR steering values ($\mathcal{G}_{c\to a}$, $\mathcal{G}_{B\to a}$, and $\mathcal{G}_{B\to c}$) as functions of the collective optomechanical coupling strength $G_j/\omega_m$ and the cavity decay rate $\kappa_j/\omega_m$. Optimal EPR steering from cavity $c$ to $a$ ($\mathcal{G}_{c\to a}$) is observed at relatively low values of $G_j/\omega_m$ and $\kappa_j/\omega_m$, specifically within the ranges $G_j/\omega_m < 3\times 10^{-3}$ and $\kappa_j/\omega_m < 3\times 10^{-3}$ for $J/\omega_m=0$ and $G_j/\omega_m < 4\times 10^{-3}$ (see \Cref{fig:fig9}(a)) and $\kappa_j/\omega_m < 3\times 10^{-3}$ for $J/\omega_m=0.1$ (see \Cref{fig:fig9}(d)) . More remarkably, as shown in \Cref{fig:fig9}(b), the one-way steering $\mathcal{G}_{B\to a}$ persists even at higher cavity decay rates. This resilience stems from the inherently directional nature of EPR steering and the robustness of collective molecular excitation. The molecular collective mode efficiently maintains its coherence and coupling to the cavity $a$, allowing quantum correlations to build up and \textit{steer} the cavity state even in the presence of significant cavity losses. Conversely, the intrinsic dissipation of the cavity hinders it from returning to the collective mode, making $\mathcal{G}_{B\to a}$ particularly robust under lossy conditions. Similarly, \Cref{fig:fig9}(c) illustrates that $\mathcal{G}_{B\to c}$ mirrors the behavior of bipartite entanglement $E_{Bc}$ (compare with \Cref{fig:fig5}(c)).

We now turn our attention to the quantum discord's dependence on coupling and decay rates. \Cref{fig:fig10} depicts density plots of $\mathcal{D}_{ca}$, $\mathcal{D}_{Ba}$, and $\mathcal{D}_{Bc}$ as a function of collective optomechanical coupling $G_j/\omega_m$ and cavity decay rate $\kappa_j/\omega_m$. When the decay rate of the cavity $\kappa_j/\omega_m$ falls below $\num{0.05}$, the two modes of the cavity exhibit strong correlations, which are predominantly mediated by the collective molecular mode. Conversely, an increase in cavity decay rate enhances the decoherence effect, inevitably weakening these inter-cavity correlations. As highlighted in \Cref{fig:fig10}(b), the quantum correlation between the molecular collective mode and the cavity mode $a$ demonstrates remarkable resilience to decoherence, persisting robustly even at high cavity decay rates when $\tilde{\Delta}_a/\omega_m = \num{0.5}$. This indicates the molecular mode's ability to maintain quantum features despite significant photon loss. Interestingly, our numerical results in \Cref{fig:fig7} reveal a consistent hierarchy in the strength of quantum discord among the subsystems: in the absence of inter-cavity coupling ($J = 0$), the discord between cavities $a$ and $c$, denoted as $\mathcal{D}_{ca}$, is generally weaker than the discord between cavity $a$ and the molecular mode $B$ ($\mathcal{D}_{Ba}$), but stronger than that between cavity $c$ and $B$ ($\mathcal{D}_{Bc}$). This ordering reflects the differing nature and strength of the direct and indirect couplings in the system. Specifically, cavity $a$ interacts directly with $B$ via a radiation-pressure-type coupling, while cavity $c$ lacks a direct connection to $a$ and interacts with $B$ only through a linear coupling. This hierarchy underscores how the structure and type of coupling influence the distribution of quantum correlations. However, when a nonzero inter-cavity coupling ($J \neq 0$) is introduced, the correlation landscape is modified: $\mathcal{D}_{Bc}$ becomes stronger than $\mathcal{D}_{Ba}$. This inversion arises because the coupling $J$ allows mode $c$ to become dynamically hybridized with mode $a$, enabling it to indirectly inherit and even amplify quantum correlations with the molecular mode $B$ particularly because its interaction with $B$ is linear and more efficient in transferring quantum information. Meanwhile, the effective coupling between $a$ and $B$ becomes weaker due to redistribution of interaction strength across the system. 
\begin{figure*}[tbh]
	\centering
	\includegraphics[width=5.5cm]{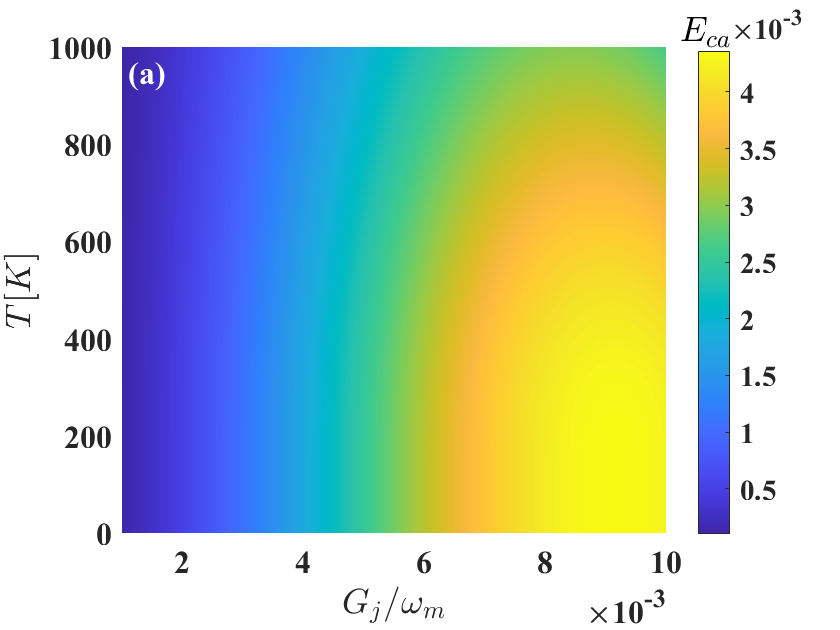}
	\includegraphics[width=5.5cm]{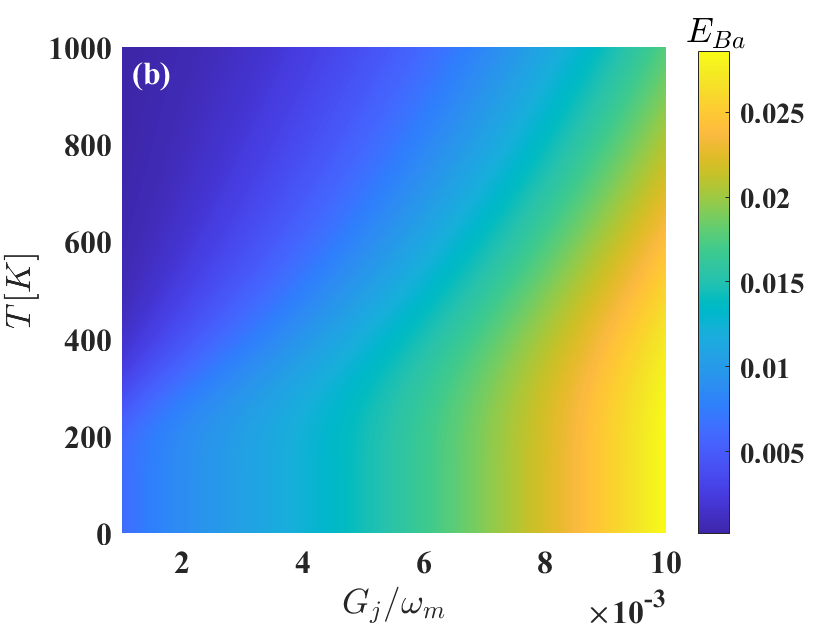}
	\includegraphics[width=5.5cm]{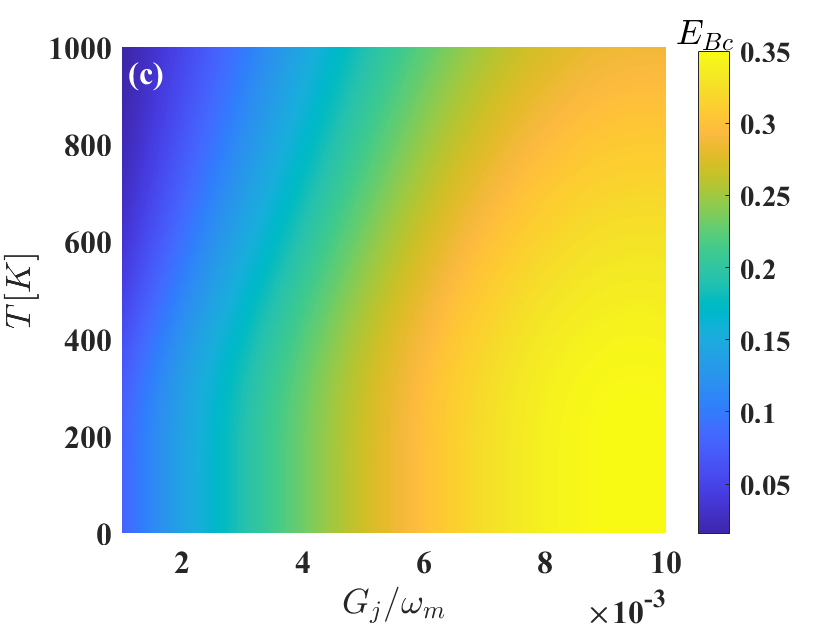}
	\includegraphics[width=5.5cm]{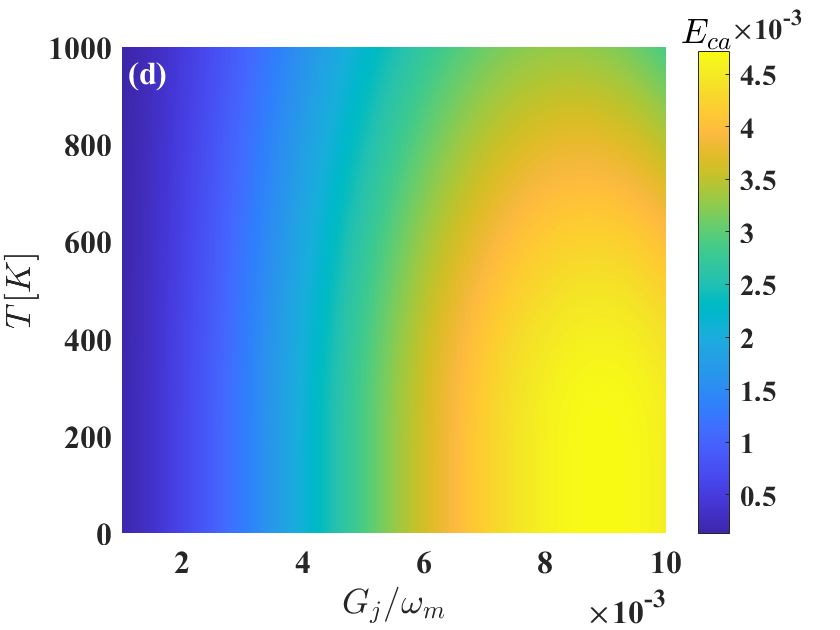}
	\includegraphics[width=5.5cm]{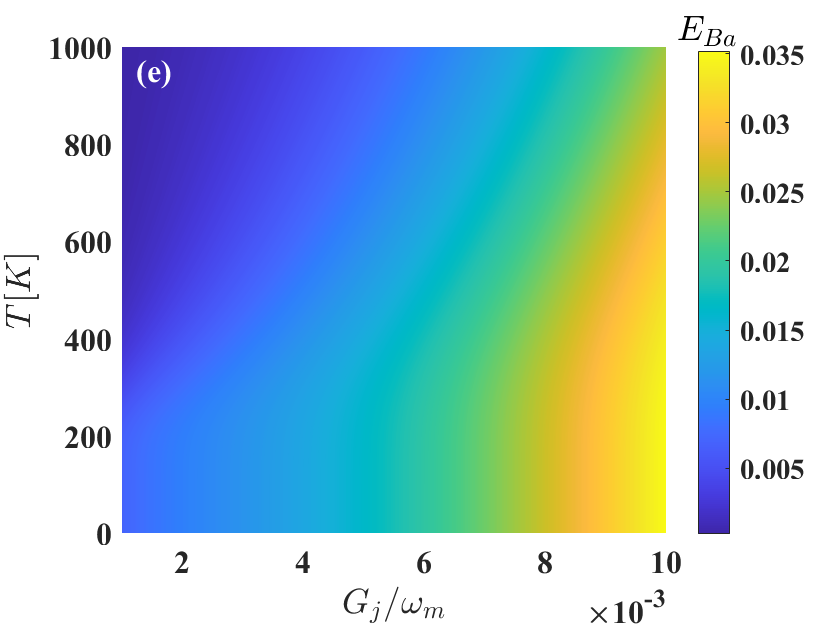}
	\includegraphics[width=5.5cm]{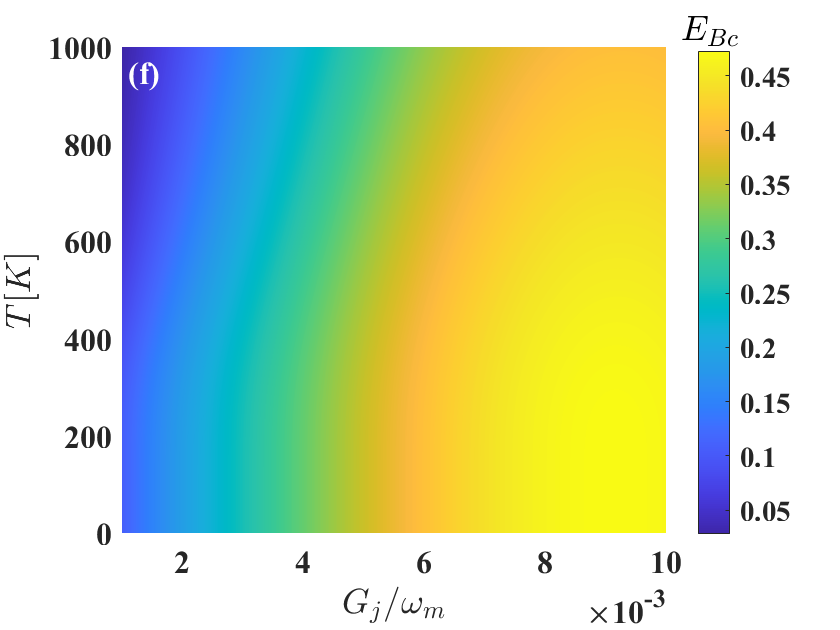}
	\caption{(a) Density plot of bipartite entanglement $E_{ca}$ between cavity modes $a$ and $c$, vs. $G_j/\omega_m$ and temperature $T$, for $\tilde{\Delta}_a/\omega_m = 1$ and ${\Delta}_c/\omega_m = -1$. (b) Entanglement $E_{Ba}$ between the molecular collective mode $B$ and cavity mode $a$, vs. $G_j/\omega_m$ and $T$, for $J/\omega_m = \num{0}$, $\tilde{\Delta}_a/\omega_m = \num{1}$ and ${\Delta}_c/\omega_m = -0.5$. (c) Entanglement $E_{Bc}$ between the molecular collective mode $B$ and cavity mode $c$, vs. $G_j/\omega_m$ and $T$, for $J/\omega_m = \num{0}$, $\tilde{\Delta}_a/\omega_m = 1$ and ${\Delta}_c/\omega_m = -1$.(d) Density plot of bipartite entanglement $E_{ca}$ between cavity modes $a$ and $c$, vs. $G_j/\omega_m$ and temperature $T$, for $J/\omega_m = \num{0}$, $\tilde{\Delta}_a/\omega_m = 1$ and ${\Delta}_c/\omega_m = -1$. (e) Entanglement $E_{Ba}$ between the molecular collective mode $B$ and cavity mode $a$, vs. $G_j/\omega_m$ and $T$, for $J/\omega_m = \num{0.1}$, $\tilde{\Delta}_a/\omega_m = \num{1}$ and ${\Delta}_c/\omega_m = -0.5$. (f) Entanglement $E_{Bc}$ between the molecular collective mode $B$ and cavity mode $c$, vs. $G_j/\omega_m$ and $T$, for $J/\omega_m = \num{0.1}$, $\tilde{\Delta}_a/\omega_m = 1$ and ${\Delta}_c/\omega_m = -1$. Common parameters for all subplots: $\mathcal{E}/\omega_m = 16$, $\gamma_m/\omega_m = \num{0.005}$, $\kappa_c/\omega_m = \num{0.0166}$, $\kappa_a/\omega_m = 1$, $g_c/\omega_m = \num{3.3e-6}$, $g_a/\omega_m = \num{2.66e-6}$, $N = \num{e6}$, and $\theta=\pi$.}
	\label{fig:fig11}
\end{figure*}

\subsection{Thermal robustness of quantum correlations}
One of the most striking and technologically significant findings of our study concerns the exceptional thermal robustness of the generated quantum correlations. \Cref{fig:fig11} displays the bipartite entanglements ($E_{ca}$, $E_{Ba}$, and $E_{Bc}$) versus temperature $T$. Remarkably, bipartite entanglement in our molecular optomechanical system exhibits exceptional resilience against thermal decoherence, persisting at temperatures up to approximately $\SI{1000}{\kelvin}$. This thermal robustness represents a significant advantage in maintaining quantum correlations above cryogenic temperatures. The fundamental physical origin of this resilience lies directly in the inherently high vibrational frequencies of the molecules. For molecular vibrations, $\hbar\omega_m$ (where $\omega_m/2\pi = \SI{30}{\tera\hertz}$) is comparable to or even exceeds the thermal energy $k_B T$ at $\SI{1000}{\kelvin}$ ($\hbar\omega_m \approx \SI{1.98e-20}{\joule}$ while $k_B T \approx \SI{1.38e-20}{\joule}$). This condition ensures that the thermal phonon occupation number ($n_{\text{th}}$) of the molecular mode remains relatively low even at high temperatures, effectively keeping the mechanical mode in a quantum-dominated regime, far less susceptible to thermal noise than the much lower frequency mechanical resonators found in conventional optomechanics.
To quantitatively illustrate this advantage, we present a comparison of typical thermal robustness in optomechanical systems.
\begin{table*}[tbh]
	\centering
	\caption{Comparative thermal robustness of optomechanical systems.}
	\label{tab:thermal_robustness}
	\begin{tabular}{lcc}
		\toprule
		\textbf{System}            & \textbf{Typical thermal robustness (K)} & \textbf{Typical vibrational frequency} \\
		\midrule
		Conventional COM \cite{Barzanjeh2021} & $< 300$                                & $\sim$MHz--GHz                               \\
		Molecular COM (McOM) (This work) & Up to 1000                             & $\sim$30GHz                                 \\
		\bottomrule
	\end{tabular}
\end{table*}
As depicted in \Cref{fig:fig11}, increasing the collective optomechanical coupling strength $G_j/\omega_m$ further enhances the entanglement across all subsystems. This is because a stronger coupling facilitates a more rapid exchange of quantum information between the optical and molecular modes, allowing the quantum interactions to outpace the rate of thermal decoherence. This effectively enables the system to counteract the detrimental effects of thermal environmental fluctuations in a more robust way.

Furthermore, we observe a distinct \textit{correlation hierarchy} in thermal robustness: $E_{Bc}$ (the entanglement between cavity $c$ and the molecular mode) is consistently more enhanced and thermally robust than $E_{ca}$ and $E_{Ba}$. The underlying physics for this differential enhancement lies in the disparate cavity decay rates: since cavity $a$ has a significantly higher decay rate ($\kappa_a/\omega_m = 1$) compared to cavity $c$ ($\kappa_c/\omega_m = \num{0.0166}$), photons in cavity $c$ have a much longer effective residence time. This extended coherent interaction period allows the collective molecular mode and cavity $c$ to establish and maintain stronger, more persistent entanglement even under adverse thermal conditions. Conversely, the rapid loss of photons from the cavity $a$ leads to faster decoherence and consequently weaker and less robust entanglement with other modes.
\section{Conclusion}\label{sec:concl}

In this work, we have investigated the generation of fundamental quantum correlations specifically entanglement, EPR steering, and quantum discord within a double-cavity molecular optomechanical system. Our findings demonstrate that these quantum correlations, significantly linking the cavity modes and the molecular collective mode, are  enhanced by increasing the collective optomechanical coupling strength. This intricate interplay between light and molecular vibrations not only enriches our understanding but also opens exciting avenues for tailoring quantum resources with unprecedented precision.

More importantly, our analysis reveals that all three quantum correlations, entanglement, EPR steering, and quantum discord are significantly enhanced when the phase difference $\theta$ of the effective optomechanical coupling satisfies the condition $\theta = m\pi$ (with $m \in \mathbb{N}$). This behavior contrasts with observations reported in previously published works, where quantum correlations were typically suppressed or not optimized at these phase values. This discrepancy highlights the distinct dynamics of our coupled cavity-molecular system and suggests that phase control may serve as a powerful tool for enhancing quantum correlations in hybrid optomechanical platforms.

An interesting aspect of our findings is the inherent thermal robustness of entanglement within these molecular subsystems. We show that entanglement persists even at temperatures near 1000 K. This thermal tolerance underscores the practical viability and profound potential of our scheme for developing robust, and room-temperature quantum technologies. This breakthrough is particularly significant for the realization of stable and scalable quantum networks, where maintaining quantum coherence at high temperatures is challenging. Our study lays a robust and comprehensive theoretical foundation, paving the way for the development of a new generation of high-performance quantum devices rooted in molecular cavity optomechanics. Our work suggests to address critical challenges in quantum engineering and paves new venues for scalable quantum information processing, quantum computing, and the practical implementation of distributed quantum networks. Furthermore, our work particularly the phase-dependent behaviour of quantum discord provides a fundamental basis for the development of ultra-sensitive gas sensors, with potential applications in environmental monitoring, medical diagnostics, and industrial safety.

\section*{Acknowledgments}
P.D. acknowledges the Iso-Lomso Fellowship at Stellenbosch Institute for Advanced Study (STIAS), Wallenberg Research Centre at Stellenbosch University, Stellenbosch 7600, South Africa, and The Institute for Advanced Study, Wissenschaftskolleg zu Berlin, Wallotstrasse 19, 14193 Berlin, Germany. This work was supported by Princess Nourah bint Abdulrahman University Researchers Supporting Project number (PNURSP2025R399), Princess Nourah bint Abdulrahman University, Riyadh, Saudi Arabia. The authors are thankful to the Deanship of Graduate Studies and Scientific Research at University of Bisha for supporting this work through the Fast-Track Research Support Program. 

\textbf{Author Contributions:} E.K.B. and P.D. conceptualized the work and carried out the simulations and analysis. D.R.K.M., R.A. and S. A.K. participated in all the discussions and provided useful methodology and suggestions for the final version of the manuscript. A.-H. A.-A. and S.G.N.E. participated in the discussions and supervised the work. All authors participated equally in the writing, discussions, and the preparation of the final version of the manuscript.

\textbf{Competing Interests:} All authors declare no competing interests.

\textbf{Data Availability:}
Relevant data are included in the manuscript and supporting information. Supplementary data are available upon reasonable request.

\bibliography{RefMolecules_QIP}
\end{document}